\documentstyle[12pt]{article}
\def\fun#1#2{\lower3.6pt\vbox{\baselineskip0pt\lineskip.9pt
  \ialign{$\mathsurround=0pt#1\hfil##\hfil$\crcr#2\crcr\sim\crcr}}}
\newcommand{\dd}{\mbox{d}}
\newcommand{\vecc}[1]{\mbox{\boldmath $#1$}}
\newcommand{\Li}{\mbox{Li}_2}

\title {Second order contributions to elastic large--angle
        Bhabha scattering}

\author {A.B~Arbuzov, E.A.~Kuraev, B.G.~Shaikhatdenov \\
{\it Bogoliubov Laboratory of Theoretical Physics, JINR},\\
 {\it Dubna, 141980, Russia} }

\date{}

\begin{document}
\maketitle

\begin{abstract}
The cross section of (quasi--)elastic large--angle
electron--positron scattering at high energies is calculated.
Radiative corrections of the orders ${\cal O}(\alpha^2L^2)$
and ${\cal O}(\alpha^2L)$,
except pure two--loop box contributions, are explicitly calculated.
In the second order we considered the following sources of corrections:
1) Virtual photonic corrections coming from
squares of 1--loop level amplitudes and their relevant interferences
(vertex--type and box--type Feynman diagrams).
2) Double soft photon emission and one--loop corrections to
single soft photon emission.
The results are presented in an analytical form.
\\[.2cm] \noindent
\end{abstract}

\section{Introduction}

The importance to know the cross section of elastic and inelastic
Bhabha scattering is emphasized mostly by experimental physics
requirements, since these processes are used for calibration.
Better than per--mille accuracy in theoretical calculations of the
cross section is required in order to obtain an
adequate level of accuracy to the present one of Standard Model parameters.
Also, gaining more insight into the properties of hadrons extracted
from experiments at colliders and meson factories, demands the above
accuracy.

Much of attention was paid to a calculation of different contributions
to the cross section to the second order of perturbation theory (PT). The
detailed analysis of contribution coming from two hard photon emission
~\cite{r1}, real and virtual pairs emission~\cite{r2} was recently
performed to the leading ($\alpha^2L^2$) and next--to--leading ($\alpha^2L$)
accuracy. In the meantime, an investigation of the 2--loop box--type
gauge--invariant set of Feynman diagrams is not completed to the
moment (their total number is 44)~\cite{r3}. Another problem is the
calculation of radiative corrections (RC) to the single hard photon
emission process. These tasks now are under close scrutiny.
As for the present paper
We summarize here all obtained in the past results concerning 2--loop
contributions to the vertex functions (Sect.~1), then calculate the
contributions coming from the squares of 1--loop Feynman amplitudes which
correspond to the both vertex-- and box--type Feynman diagrams (Sect.~2).
In Sect.~3 we give the expressions of contributions
of two soft photons emission processes.
In this paper we do not consider the effects of vacuum polarization
inserted into the virtual photon Green function since it was examined earlier
in paper~\cite{r2}. Our final result is given in~Eq.(\ref{eq:21}).

\section{2--loop vertex contribution}

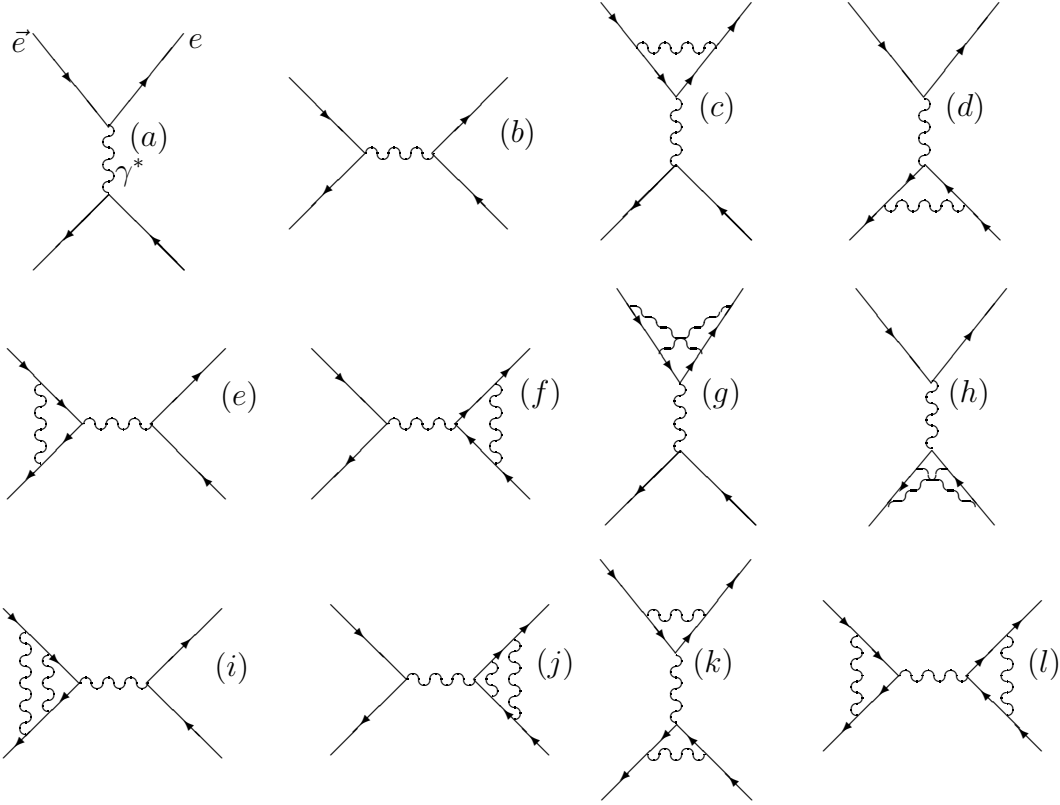
\begin{figure}[h]
\unitlength=0.50mm
\special{em:linewidth 0.4pt}
\linethickness{0.4pt}
\begin{picture}(286.81,216.00)
\put(189.00,174.00){\oval(3.00,3.00)[l]}
\put(189.00,177.00){\oval(3.00,3.00)[r]}
\put(189.00,180.00){\oval(3.00,3.00)[l]}
\put(189.00,183.00){\oval(3.00,3.00)[r]}
\put(189.00,186.00){\oval(3.00,3.00)[l]}
\put(189.00,189.00){\oval(3.00,3.00)[r]}
\put(169.00,216.00){\line(4,-5){20.00}}
\put(189.00,191.00){\line(4,5){20.00}}
\put(189.00,173.00){\line(-1,-1){20.00}}
\put(189.00,173.00){\line(1,-1){20.00}}
\put(189.00,173.00){\vector(-1,-1){12.00}}
\put(209.00,153.00){\vector(-1,1){9.00}}
\put(180.00,204.00){\oval(3.00,3.00)[t]}
\put(183.00,204.00){\oval(3.00,3.00)[b]}
\put(186.00,204.00){\oval(3.00,3.00)[t]}
\put(189.00,204.00){\oval(3.00,3.00)[b]}
\put(192.00,204.00){\oval(3.00,3.00)[t]}
\put(195.00,204.00){\oval(3.00,3.00)[b]}
\put(198.00,204.00){\oval(3.00,3.00)[t]}
\put(199.81,188.00){\makebox(0,0)[cc]{$(c)$}}
\put(171.80,212.67){\vector(3,-4){1.67}}
\put(184.80,196.67){\vector(3,-4){1.67}}
\put(234.81,216.00){\line(4,-5){20.00}}
\put(254.81,191.00){\line(4,5){20.00}}
\put(255.14,173.00){\line(-1,-1){20.00}}
\put(255.14,173.00){\line(1,-1){20.00}}
\put(265.81,188.00){\makebox(0,0)[cc]{$(d)$}}
\
\put(203.47,208.67){\vector(3,4){1.67}}
\put(192.14,194.67){\vector(3,4){1.67}}
\
\put(18.00,208.00){\line(4,-5){20.00}}
\put(38.00,183.00){\line(4,5){20.00}}
\put(38.14,165.00){\line(-1,-1){20.00}}
\put(38.14,165.00){\line(1,-1){20.00}}
\put(38.14,165.00){\vector(-1,-1){12.00}}
\put(58.14,145.00){\vector(-1,1){9.00}}
\put(47.00,194.25){\vector(3,4){1.67}}
\put(48.81,180.00){\makebox(0,0)[cc]{$(a)$}}
\put(43.81,171.00){\makebox(0,0)[cc]{$\gamma^{*}$}}
\put(14.14,206.00){\makebox(0,0)[cc]{$\vec{e}$}}
\put(61.14,206.00){\makebox(0,0)[cc]{$e$}}
\
\put(86.14,196.00){\line(1,-1){20.00}}
\put(106.14,176.00){\line(-1,-1){20.00}}
\put(144.14,196.00){\line(-1,-1){20.00}}
\put(124.14,176.00){\line(1,-1){20.00}}
\put(11.14,124.00){\line(1,-1){20.00}}
\put(31.14,104.00){\line(-1,-1){20.00}}
\put(69.14,124.00){\line(-1,-1){20.00}}
\put(49.14,104.00){\line(1,-1){20.00}}
\put(92.14,124.00){\line(1,-1){20.00}}
\put(112.14,104.00){\line(-1,-1){20.00}}
\put(150.14,124.00){\line(-1,-1){20.00}}
\put(130.14,104.00){\line(1,-1){20.00}}
\put(190.14,97.00){\line(-1,-1){20.00}}
\put(190.14,97.00){\line(1,-1){20.00}}
\put(190.14,97.00){\vector(-1,-1){12.00}}
\put(210.14,77.00){\vector(-1,1){9.00}}
\put(200.81,112.00){\makebox(0,0)[cc]{$(g)$}}
\put(236.81,140.00){\line(4,-5){20.00}}
\put(256.81,115.00){\line(4,5){20.00}}
\put(266.81,112.00){\makebox(0,0)[cc]{$(h)$}}
\put(10.14,55.00){\line(1,-1){20.00}}
\put(30.14,35.00){\line(-1,-1){20.00}}
\put(68.14,55.00){\line(-1,-1){20.00}}
\put(48.14,35.00){\line(1,-1){20.00}}
\put(97.14,56.00){\line(1,-1){20.00}}
\put(117.14,36.00){\line(-1,-1){20.00}}
\put(155.14,56.00){\line(-1,-1){20.00}}
\put(135.14,36.00){\line(1,-1){20.00}}
\put(168.81,68.00){\line(4,-5){20.00}}
\put(188.81,43.00){\line(4,5){20.00}}
\put(189.14,24.00){\line(-1,-1){20.00}}
\put(189.14,24.00){\line(1,-1){20.00}}
\put(198.81,40.00){\makebox(0,0)[cc]{$(k)$}}
\put(171.47,64.67){\vector(3,-4){1.67}}
\put(184.47,48.67){\vector(3,-4){1.67}}
\put(203.14,60.67){\vector(3,4){1.67}}
\put(191.47,46.34){\vector(3,4){1.67}}
\put(228.14,57.00){\line(1,-1){20.00}}
\put(248.14,37.00){\line(-1,-1){20.00}}
\put(286.14,57.00){\line(-1,-1){20.00}}
\put(266.14,37.00){\line(1,-1){20.00}}
\put(146.81,181.00){\makebox(0,0)[cc]{$(b)$}}
\put(152.81,112.00){\makebox(0,0)[cc]{$(f)$}}
\put(72.81,111.00){\makebox(0,0)[cc]{$(e)$}}
\put(156.81,40.00){\makebox(0,0)[cc]{$(j)$}}
\put(70.81,39.00){\makebox(0,0)[cc]{$(i)$}}
\put(286.81,40.00){\makebox(0,0)[cc]{$(l)$}}
\
\put(136.00,164.00){\vector(-1,1){1.03}}
\put(132.00,184.00){\vector(1,1){2.17}}
\put(97.00,167.00){\vector(-1,-1){2.06}}
\put(16.00,119.00){\vector(1,-1){0.94}}
\put(26.00,109.00){\vector(1,-1){1.01}}
\put(27.00,100.00){\vector(-1,-1){1.08}}
\put(18.00,91.00){\vector(-1,-1){1.98}}
\put(64.00,89.00){\vector(-1,1){1.08}}
\put(62.00,117.00){\vector(1,1){0.92}}
\put(100.00,116.00){\vector(1,-1){1.04}}
\put(102.00,94.00){\vector(-1,-1){0.96}}
\put(133.00,107.00){\vector(1,1){0.97}}
\put(134.00,100.00){\vector(-1,1){1.01}}
\put(144.00,118.00){\vector(1,1){0.95}}
\put(144.00,90.00){\vector(-1,1){1.05}}
\put(186.00,21.00){\vector(-1,-1){1.06}}
\put(175.00,10.00){\vector(-1,-1){0.99}}
\put(193.00,20.00){\vector(-1,1){0.99}}
\put(204.00,9.00){\vector(-1,1){1.06}}
\put(13.00,52.00){\vector(1,-1){0.99}}
\put(24.00,41.00){\vector(1,-1){0.96}}
\put(26.00,31.00){\vector(-1,-1){1.04}}
\put(14.00,19.00){\vector(-1,-1){0.97}}
\put(59.00,46.00){\vector(1,1){0.94}}
\put(59.00,24.00){\vector(-1,1){0.97}}
\put(105.00,48.00){\vector(1,-1){1.02}}
\put(106.00,25.00){\vector(-1,-1){1.03}}
\put(149.00,50.00){\vector(1,1){1.00}}
\put(151.00,20.00){\vector(-1,1){1.00}}
\put(234.00,51.00){\vector(1,-1){1.08}}
\put(243.00,42.00){\vector(1,-1){1.03}}
\put(244.00,33.00){\vector(-1,-1){1.07}}
\put(234.00,23.00){\vector(-1,-1){1.02}}
\put(270.00,41.00){\vector(1,1){1.07}}
\put(280.00,51.00){\vector(1,1){1.02}}
\put(281.00,22.00){\vector(-1,1){0.98}}
\put(271.00,32.00){\vector(-1,1){1.03}}
\put(255.00,174.00){\oval(3.00,3.00)[l]}
\put(255.00,177.00){\oval(3.00,3.00)[r]}
\put(255.00,180.00){\oval(3.00,3.00)[l]}
\put(255.00,183.00){\oval(3.00,3.00)[r]}
\put(255.00,186.00){\oval(3.00,3.00)[l]}
\put(255.00,189.00){\oval(3.00,3.00)[r]}
\put(38.00,167.00){\oval(3.00,3.00)[l]}
\put(38.00,170.00){\oval(3.00,3.00)[r]}
\put(38.00,173.00){\oval(3.00,3.00)[l]}
\put(38.00,176.00){\oval(3.00,3.00)[r]}
\put(38.00,179.00){\oval(3.00,3.00)[l]}
\put(38.00,182.00){\oval(3.00,3.00)[r]}
\put(20.00,95.00){\oval(3.00,3.00)[l]}
\put(20.00,98.00){\oval(3.00,3.00)[r]}
\put(20.00,101.00){\oval(3.00,3.00)[l]}
\put(20.00,104.00){\oval(3.00,3.00)[r]}
\put(20.00,107.00){\oval(3.00,3.00)[l]}
\put(20.00,110.00){\oval(3.00,3.00)[r]}
\put(141.00,95.00){\oval(3.00,3.00)[r]}
\put(141.00,98.00){\oval(3.00,3.00)[l]}
\put(141.00,101.00){\oval(3.00,3.00)[r]}
\put(141.00,104.00){\oval(3.00,3.00)[l]}
\put(141.00,107.00){\oval(3.00,3.00)[r]}
\put(22.00,29.00){\oval(3.00,3.00)[l]}
\put(22.00,32.00){\oval(3.00,3.00)[r]}
\put(22.00,35.00){\oval(3.00,3.00)[l]}
\put(22.00,38.00){\oval(3.00,3.00)[r]}
\put(22.00,41.00){\oval(3.00,3.00)[l]}
\put(16.00,35.00){\oval(3.00,3.00)[l]}
\put(16.00,38.00){\oval(3.00,3.00)[r]}
\put(16.00,41.00){\oval(3.00,3.00)[l]}
\put(16.00,44.00){\oval(3.00,3.00)[r]}
\put(16.00,47.00){\oval(3.00,3.00)[l]}
\put(16.00,32.00){\oval(3.00,3.00)[r]}
\put(16.00,29.00){\oval(3.00,3.00)[l]}
\put(16.00,26.00){\oval(3.00,3.00)[r]}
\put(16.00,23.00){\oval(3.00,3.00)[l]}
\put(140.00,33.00){\oval(3.00,3.00)[r]}
\put(140.00,36.00){\oval(3.00,3.00)[l]}
\put(140.00,39.00){\oval(3.00,3.00)[r]}
\put(146.00,36.00){\oval(3.00,3.00)[l]}
\put(146.00,39.00){\oval(3.00,3.00)[r]}
\put(146.00,42.00){\oval(3.00,3.00)[l]}
\put(146.00,45.00){\oval(3.00,3.00)[r]}
\put(146.00,33.00){\oval(3.00,3.00)[r]}
\put(146.00,30.00){\oval(3.00,3.00)[l]}
\put(146.00,27.00){\oval(3.00,3.00)[r]}
\put(144.00,27.00){\vector(-1,1){0.93}}
\put(190.00,98.00){\oval(3.00,3.00)[l]}
\put(190.00,101.00){\oval(3.00,3.00)[r]}
\put(190.00,104.00){\oval(3.00,3.00)[l]}
\put(190.00,107.00){\oval(3.00,3.00)[r]}
\put(190.00,110.00){\oval(3.00,3.00)[l]}
\put(190.00,113.00){\oval(3.00,3.00)[r]}
\put(257.00,99.00){\oval(3.00,3.00)[l]}
\put(257.00,102.00){\oval(3.00,3.00)[r]}
\put(257.00,105.00){\oval(3.00,3.00)[l]}
\put(257.00,108.00){\oval(3.00,3.00)[r]}
\put(257.00,111.00){\oval(3.00,3.00)[l]}
\put(257.00,114.00){\oval(3.00,3.00)[r]}
\put(189.00,32.00){\oval(3.00,3.00)[l]}
\put(189.00,35.00){\oval(3.00,3.00)[r]}
\put(189.00,38.00){\oval(3.00,3.00)[l]}
\put(189.00,41.00){\oval(3.00,3.00)[r]}
\put(189.00,29.00){\oval(3.00,3.00)[r]}
\put(189.00,26.00){\oval(3.00,3.00)[l]}
\put(237.00,34.00){\oval(3.00,3.00)[l]}
\put(237.00,37.00){\oval(3.00,3.00)[r]}
\put(237.00,40.00){\oval(3.00,3.00)[l]}
\put(237.00,43.00){\oval(3.00,3.00)[r]}
\put(237.00,31.00){\oval(3.00,3.00)[r]}
\put(237.00,28.00){\oval(3.00,3.00)[l]}
\put(237.00,46.00){\oval(3.00,3.00)[l]}
\put(277.00,31.00){\oval(3.00,3.00)[l]}
\put(277.00,34.00){\oval(3.00,3.00)[r]}
\put(277.00,37.00){\oval(3.00,3.00)[l]}
\put(277.00,40.00){\oval(3.00,3.00)[r]}
\put(277.00,28.00){\oval(3.00,3.00)[r]}
\put(277.00,43.00){\oval(3.00,3.00)[l]}
\put(277.00,46.00){\oval(3.00,3.00)[r]}
\put(142.00,43.00){\vector(1,1){1.01}}
\put(20.00,113.00){\oval(3.00,3.00)[l]}
\put(249.00,162.00){\oval(3.00,3.00)[t]}
\put(252.00,162.00){\oval(3.00,3.00)[b]}
\put(255.00,162.00){\oval(3.00,3.00)[t]}
\put(258.00,162.00){\oval(3.00,3.00)[b]}
\put(261.00,162.00){\oval(3.00,3.00)[t]}
\put(264.00,162.00){\oval(3.00,3.00)[b]}
\put(246.00,162.00){\oval(3.00,3.00)[b]}
\put(271.00,157.00){\vector(-1,1){0.97}}
\put(261.00,167.00){\vector(-1,1){0.95}}
\put(251.00,169.00){\vector(-1,-1){1.04}}
\put(108.00,176.00){\oval(3.00,3.00)[t]}
\put(111.00,176.00){\oval(3.00,3.00)[b]}
\put(114.00,176.00){\oval(3.00,3.00)[t]}
\put(117.00,176.00){\oval(3.00,3.00)[b]}
\put(120.00,176.00){\oval(3.00,3.00)[t]}
\put(123.00,176.00){\oval(3.00,3.00)[b]}
\put(95.00,187.00){\vector(1,-1){2.05}}
\put(26.00,198.00){\vector(2,-3){2.14}}
\put(114.00,104.00){\oval(3.00,3.00)[t]}
\put(117.00,104.00){\oval(3.00,3.00)[b]}
\put(120.00,104.00){\oval(3.00,3.00)[t]}
\put(123.00,104.00){\oval(3.00,3.00)[b]}
\put(126.00,104.00){\oval(3.00,3.00)[t]}
\put(129.00,104.00){\oval(3.00,3.00)[b]}
\put(33.00,104.00){\oval(3.00,3.00)[t]}
\put(36.00,104.00){\oval(3.00,3.00)[b]}
\put(39.00,104.00){\oval(3.00,3.00)[t]}
\put(42.00,104.00){\oval(3.00,3.00)[b]}
\put(45.00,104.00){\oval(3.00,3.00)[t]}
\put(48.00,104.00){\oval(3.00,3.00)[b]}
\put(32.00,35.00){\oval(3.00,3.00)[t]}
\put(35.00,35.00){\oval(3.00,3.00)[b]}
\put(38.00,35.00){\oval(3.00,3.00)[t]}
\put(41.00,35.00){\oval(3.00,3.00)[b]}
\put(44.00,35.00){\oval(3.00,3.00)[t]}
\put(47.00,35.00){\oval(3.00,3.00)[b]}
\put(119.00,36.00){\oval(3.00,3.00)[t]}
\put(122.00,36.00){\oval(3.00,3.00)[b]}
\put(125.00,36.00){\oval(3.00,3.00)[t]}
\put(128.00,36.00){\oval(3.00,3.00)[b]}
\put(131.00,36.00){\oval(3.00,3.00)[t]}
\put(134.00,36.00){\oval(3.00,3.00)[b]}
\put(183.00,53.00){\oval(3.00,3.00)[t]}
\put(186.00,53.00){\oval(3.00,3.00)[b]}
\put(189.00,53.00){\oval(3.00,3.00)[t]}
\put(192.00,53.00){\oval(3.00,3.00)[b]}
\put(195.00,53.00){\oval(3.00,3.00)[t]}
\put(183.00,16.00){\oval(3.00,3.00)[b]}
\put(186.00,16.00){\oval(3.00,3.00)[t]}
\put(189.00,16.00){\oval(3.00,3.00)[b]}
\put(192.00,16.00){\oval(3.00,3.00)[t]}
\put(195.00,16.00){\oval(3.00,3.00)[b]}
\put(250.00,37.00){\oval(3.00,3.00)[t]}
\put(253.00,37.00){\oval(3.00,3.00)[b]}
\put(256.00,37.00){\oval(3.00,3.00)[t]}
\put(259.00,37.00){\oval(3.00,3.00)[b]}
\put(262.00,37.00){\oval(3.00,3.00)[t]}
\put(265.00,37.00){\oval(3.00,3.00)[b]}
\put(240.00,158.00){\vector(-1,-1){1.01}}
\put(141.00,110.00){\oval(3.00,3.00)[l]}
\put(141.00,113.00){\oval(3.00,3.00)[r]}
\
\put(265.00,83.00){\oval(7.00,1.00)[rt]}
\put(265.00,84.80){\oval(2.00,1.00)[lb]}
\put(261.00,85.80){\oval(6.00,1.00)[rt]}
\put(261.00,87.60){\oval(2.00,1.00)[lb]}
\put(257.00,88.60){\oval(6.00,1.00)[rt]}
\put(257.00,90.50){\oval(2.00,1.00)[lb]}
\put(253.00,91.50){\oval(6.00,1.00)[rt]}
\
\put(249.00,83.00){\oval(7.00,1.00)[lt]}
\put(249.00,84.80){\oval(2.00,1.00)[rb]}
\put(253.00,85.80){\oval(6.00,1.00)[lt]}
\put(253.00,87.60){\oval(2.00,1.00)[rb]}
\put(257.00,88.60){\oval(6.00,1.00)[lt]}
\put(257.00,90.50){\oval(2.00,1.00)[rb]}
\put(261.00,91.50){\oval(6.00,1.00)[lt]}
\
\
\put(187.50,123.50){\oval(6.00,1.00)[lt]}
\put(187.50,125.30){\oval(2.00,1.00)[rb]}
\put(191.50,126.30){\oval(6.00,1.00)[lt]}
\put(191.50,128.10){\oval(2.00,1.00)[rb]}
\put(195.50,129.10){\oval(6.00,1.00)[lt]}
\put(195.50,130.90){\oval(2.00,1.00)[rb]}
\put(199.50,131.90){\oval(6.00,1.00)[lt]}
\put(199.50,133.70){\oval(2.00,1.00)[rb]}
\put(203.50,134.70){\oval(5.00,1.00)[lt]}
\
\put(192.80,123.50){\oval(6.00,1.00)[rt]}
\put(192.80,125.30){\oval(2.00,1.00)[lb]}
\put(188.80,126.30){\oval(6.00,1.00)[rt]}
\put(188.80,128.10){\oval(2.00,1.00)[lb]}
\put(184.80,129.10){\oval(6.00,1.00)[rt]}
\put(184.80,130.90){\oval(2.00,1.00)[lb]}
\put(180.80,131.90){\oval(6.00,1.00)[rt]}
\put(180.80,133.80){\oval(2.00,1.00)[lb]}
\put(176.80,134.80){\oval(6.00,1.00)[rt]}
\put(257.00,97.00){\line(5,-6){16.63}}
\put(257.00,97.00){\line(-5,-6){16.75}}
\put(190.00,115.00){\line(2,3){16.78}}
\put(190.00,115.00){\line(-2,3){16.66}}
\put(192.00,118.00){\vector(2,3){1.94}}
\put(178.00,133.00){\vector(2,-3){1.90}}
\put(186.00,121.00){\vector(2,-3){1.96}}
\put(198.00,127.00){\vector(2,3){1.92}}
\put(266.00,86.00){\vector(-3,4){3.03}}
\put(252.00,91.00){\vector(-3,-4){3.76}}
\put(265.00,125.00){\vector(3,4){2.96}}
\put(244.00,131.00){\vector(3,-4){3.04}}
\put(263.00,201.00){\vector(3,4){2.99}}
\put(243.00,206.00){\vector(2,-3){2.08}}
\end{picture}
\caption{Vertex diagrams up to 2--loop level.}
\end{figure}

The corresponding Feynman diagrams up to 2--loop level are depicted
in Fig.~1 (there are four more diagrams coming from crossing channels
to Fig.~1(g,h,i,j)).
We use the following asymptotes of the fermion vertex function in the case
of space--like and time--like 4--vectors of virtual photons~\cite{r4}:
\begin{eqnarray}
\Gamma_{\mu}(q^2)&=&\gamma_{\mu} \left[ 1+\frac{\alpha}{\pi}
\Gamma^{(2)}(q^2)+\left(\frac{\alpha}{\pi}\right)^2\Gamma^{(4)}(q^2)
\right],\quad{\mathrm{where}}\quad q^2=s>0\quad{\mathrm{or}}\quad q^2=t<0\nonumber \\
\Gamma^{(2)}(s)&=&\frac{L-1}{2}L_{\lambda}-\frac{1}{4}L^2+\frac{3}{4}L+
\frac{\pi^2}{3}-1+\mathrm{i}\pi\left(\frac{1}{2}L-\frac{1}{2}L_{\lambda}
-\frac{3}{4}\right), \nonumber \\
\Gamma^{(2)}(t)&=&(L_t-1)\left(\frac{1}{2}L_{\lambda}+1\right)
-\frac{1}{4}L_t^2-\frac{1}{4}L_t+\frac{\pi^2}{12}\, , \nonumber \\
\mathrm{Re}\; \Gamma^{(4)}(s)&=&\frac{1}{8}(L^2-2L+1-\pi^2)L_{\lambda}^2
+\frac{1}{2}L_{\lambda}\biggl[(L-1)\left(-\frac{1}{4}L^2+\frac{3}{4}L-1
+\frac{\pi^2}{3}\right)+ \pi^2\biggl(\frac{1}{2}L  \nonumber \\
&-& \frac{3}{4}\biggr)\biggr] + \frac{1}{32}L^4
-\frac{3}{16}L^3+\left(\frac{17}{32}-\frac{5\pi^2}{24}\right)L^2
+\left(-\frac{21}{32}+\frac{3}{2}\zeta(3)+\frac{17\pi^2}{36}\right)L
+{\cal O}(1), \nonumber \\
\Gamma^{(4)}(t)&=&\frac{1}{32}L^4_t-\frac{3}{16}L_t^3
+\left(\frac{17}{32}-\frac{\pi^2}{48}\right)L_t^2+\left(-\frac{21}{32}
-\frac{\pi^2}{16}+\frac{3}{2}\zeta(3)\right)L_t+\frac{1}{8}L_{\lambda}^2
(L_t-1)^2 \nonumber \\
&+&\frac{1}{2}L_{\lambda}(L_t-1)\left(-\frac{1}{4}L_t^2+\frac{3}{4}L_t-1
+\frac{\pi^2}{12}\right)+{\cal O}(1), \\
&&L=\ln\frac{s}{m^2}\, , \qquad L_t=\frac{-t}{m^2}\, ,
\qquad L_{\lambda}=\ln\frac{\lambda^2}{m^2}\, , \qquad
\zeta(3) \approx 1.2020569\, . \nonumber
\end{eqnarray}
In these formul\ae we have retained only the Dirac form factor of electron and
dropped the Pauli one, since its contribution is suppressed
by the factor of $m^2/s$.

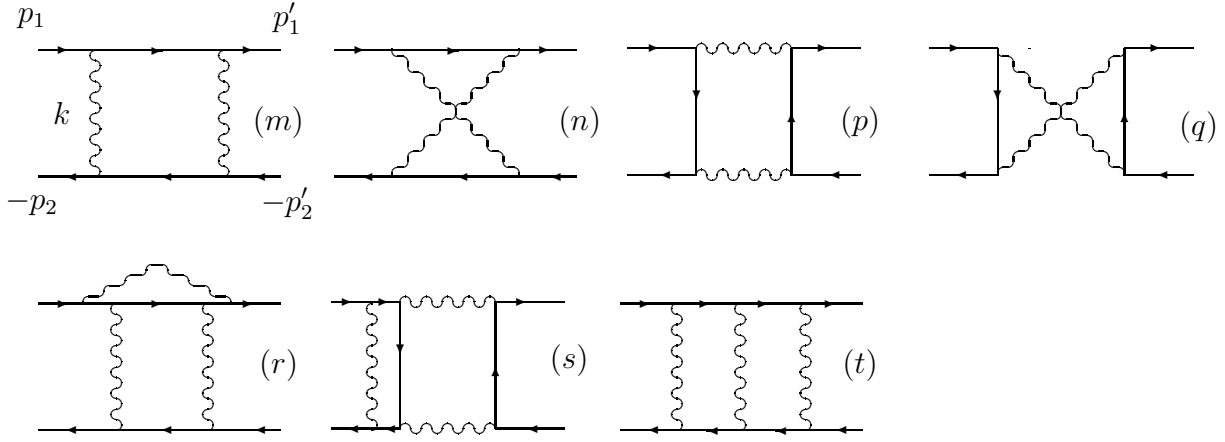
\begin{figure}[h]
\unitlength=2.00pt
\special{em:linewidth 0.4pt}
\linethickness{0.4pt}
\begin{picture}(230.00,86.67)
\put(21.33,57.67){\oval(2.00,2.00)[r]}
\put(21.33,59.67){\oval(2.00,2.00)[l]}
\put(21.33,61.67){\oval(2.00,2.00)[r]}
\put(21.33,63.67){\oval(2.00,2.00)[l]}
\put(21.33,65.67){\oval(2.00,2.00)[r]}
\put(21.33,67.67){\oval(2.00,2.00)[l]}
\put(21.33,69.67){\oval(2.00,2.00)[r]}
\put(21.33,71.67){\oval(2.00,2.00)[l]}
\put(21.33,73.67){\oval(2.00,2.00)[r]}
\put(21.33,75.67){\oval(2.00,2.00)[l]}
\put(21.33,77.67){\oval(2.00,2.00)[r]}
\put(21.33,79.67){\oval(2.00,2.00)[l]}
\put(45.66,57.67){\oval(2.00,2.00)[r]}
\put(45.66,59.67){\oval(2.00,2.00)[l]}
\put(45.66,61.67){\oval(2.00,2.00)[r]}
\put(45.66,63.67){\oval(2.00,2.00)[l]}
\put(45.66,65.67){\oval(2.00,2.00)[r]}
\put(45.66,67.67){\oval(2.00,2.00)[l]}
\put(45.66,69.67){\oval(2.00,2.00)[r]}
\put(45.66,71.67){\oval(2.00,2.00)[l]}
\put(45.66,73.67){\oval(2.00,2.00)[r]}
\put(45.66,75.67){\oval(2.00,2.00)[l]}
\put(45.66,77.67){\oval(2.00,2.00)[r]}
\put(45.66,79.67){\oval(2.00,2.00)[l]}
\put(10.66,80.67){\line(1,0){45.67}}
\put(10.66,56.67){\line(1,0){45.67}}
\put(15.33,80.67){\vector(1,0){1.00}}
\put(50.33,80.67){\vector(1,0){1.00}}
\put(17.00,56.67){\vector(-1,0){1.33}}
\put(35.00,56.67){\vector(-1,0){1.33}}
\put(52.67,56.67){\vector(-1,0){1.33}}
\put(9.33,86.67){\makebox(0,0)[cc]{$p_1$}}
\put(57.66,86.67){\makebox(0,0)[cc]{$p_1'$}}
\put(57.66,51.67){\makebox(0,0)[cc]{$-p_2'$}}
\put(9.33,51.67){\makebox(0,0)[cc]{$-p_2$}}
\put(15.00,68.67){\makebox(0,0)[cc]{$k$}}
\put(66.66,80.67){\line(1,0){45.67}}
\put(66.66,56.67){\line(1,0){45.67}}
\put(71.33,80.67){\vector(1,0){1.00}}
\put(106.33,80.67){\vector(1,0){1.00}}
\put(73.00,56.67){\vector(-1,0){1.33}}
\put(91.00,56.67){\vector(-1,0){1.33}}
\put(108.67,56.67){\vector(-1,0){1.33}}
\put(135.00,81.00){\line(0,-1){0.08}}
\put(153.00,81.00){\line(0,-1){0.08}}
\put(136.00,81.00){\oval(2.00,2.00)[t]}
\put(138.00,81.00){\oval(2.00,2.00)[b]}
\put(140.00,81.00){\oval(2.00,2.00)[t]}
\put(142.00,81.00){\oval(2.00,2.00)[b]}
\put(144.00,81.00){\oval(2.00,2.00)[t]}
\put(146.00,81.00){\oval(2.00,2.00)[b]}
\put(148.00,81.00){\oval(2.00,2.00)[t]}
\put(150.00,81.00){\oval(2.00,2.00)[b]}
\put(152.00,81.00){\oval(2.00,2.00)[t]}
\put(136.00,57.00){\oval(2.00,2.00)[t]}
\put(138.00,57.00){\oval(2.00,2.00)[b]}
\put(140.00,57.00){\oval(2.00,2.00)[t]}
\put(142.00,57.00){\oval(2.00,2.00)[b]}
\put(144.00,57.00){\oval(2.00,2.00)[t]}
\put(146.00,57.00){\oval(2.00,2.00)[b]}
\put(148.00,57.00){\oval(2.00,2.00)[t]}
\put(150.00,57.00){\oval(2.00,2.00)[b]}
\put(152.00,57.00){\oval(2.00,2.00)[t]}
\put(198.00,81.00){\line(0,-1){0.08}}
\put(216.00,81.00){\line(0,-1){0.08}}
\put(135.00,81.00){\line(0,-1){0.01}}
\put(153.00,81.00){\line(0,-1){0.01}}
\put(135.00,57.00){\line(0,0){0.00}}
\put(135.00,57.00){\line(0,0){0.00}}
\put(79.00,57.00){\oval(3.00,3.00)[lt]}
\put(79.00,60.00){\oval(3.00,3.00)[rb]}
\put(82.00,60.00){\oval(3.00,3.00)[lt]}
\put(82.00,63.00){\oval(3.00,3.00)[rb]}
\put(85.00,63.00){\oval(3.00,3.00)[lt]}
\put(85.00,66.00){\oval(3.00,3.00)[rb]}
\put(88.00,66.00){\oval(3.00,3.00)[lt]}
\put(88.00,69.00){\oval(3.00,3.00)[rb]}
\put(91.00,69.00){\oval(3.00,3.00)[lt]}
\put(91.00,72.00){\oval(3.00,3.00)[rb]}
\put(94.00,72.00){\oval(3.00,3.00)[lt]}
\put(94.00,75.00){\oval(3.00,3.00)[rb]}
\put(97.00,75.00){\oval(3.00,3.00)[lt]}
\put(97.00,78.00){\oval(3.00,3.00)[rb]}
\put(100.00,78.00){\oval(3.00,3.00)[lt]}
\put(100.00,81.00){\oval(3.00,3.00)[rb]}
\
\put(100.00,57.00){\oval(3.00,3.00)[rt]}
\put(100.00,60.00){\oval(3.00,3.00)[lb]}
\put(97.00,60.00){\oval(3.00,3.00)[rt]}
\put(97.00,63.00){\oval(3.00,3.00)[lb]}
\put(94.00,63.00){\oval(3.00,3.00)[rt]}
\put(94.00,66.00){\oval(3.00,3.00)[lb]}
\put(91.00,66.00){\oval(3.00,3.00)[rt]}
\put(91.00,69.00){\oval(3.00,3.00)[lb]}
\put(88.00,69.00){\oval(3.00,3.00)[rt]}
\put(88.00,72.00){\oval(3.00,3.00)[lb]}
\put(85.00,72.00){\oval(3.00,3.00)[rt]}
\put(85.00,75.00){\oval(3.00,3.00)[lb]}
\put(82.00,75.00){\oval(3.00,3.00)[rt]}
\put(82.00,78.00){\oval(3.00,3.00)[lb]}
\put(79.00,78.00){\oval(3.00,3.00)[rt]}
\put(79.00,81.00){\oval(3.00,3.00)[lb]}
\
\put(193.50,57.00){\oval(3.00,3.00)[lt]}
\put(193.50,60.00){\oval(3.00,3.00)[rb]}
\put(196.50,60.00){\oval(3.00,3.00)[lt]}
\put(196.50,63.00){\oval(3.00,3.00)[rb]}
\put(199.50,63.00){\oval(3.00,3.00)[lt]}
\put(199.50,66.00){\oval(3.00,3.00)[rb]}
\put(202.50,66.00){\oval(3.00,3.00)[lt]}
\put(202.50,69.00){\oval(3.00,3.00)[rb]}
\put(205.50,69.00){\oval(3.00,3.00)[lt]}
\put(205.50,72.00){\oval(3.00,3.00)[rb]}
\put(208.50,72.00){\oval(3.00,3.00)[lt]}
\put(208.50,75.00){\oval(3.00,3.00)[rb]}
\put(211.50,75.00){\oval(3.00,3.00)[lt]}
\put(211.50,78.00){\oval(3.00,3.00)[rb]}
\put(214.50,78.00){\oval(3.00,3.00)[lt]}
\put(214.50,81.00){\oval(3.00,3.00)[rb]}
\
\put(214.50,57.00){\oval(3.00,3.00)[rt]}
\put(214.50,60.00){\oval(3.00,3.00)[lb]}
\put(211.50,60.00){\oval(3.00,3.00)[rt]}
\put(211.50,63.00){\oval(3.00,3.00)[lb]}
\put(208.50,63.00){\oval(3.00,3.00)[rt]}
\put(208.50,66.00){\oval(3.00,3.00)[lb]}
\put(205.50,66.00){\oval(3.00,3.00)[rt]}
\put(205.50,69.00){\oval(3.00,3.00)[lb]}
\put(202.50,69.00){\oval(3.00,3.00)[rt]}
\put(202.50,72.00){\oval(3.00,3.00)[lb]}
\put(199.50,72.00){\oval(3.00,3.00)[rt]}
\put(199.50,75.00){\oval(3.00,3.00)[lb]}
\put(196.50,75.00){\oval(3.00,3.00)[rt]}
\put(196.50,78.00){\oval(3.00,3.00)[lb]}
\put(193.50,78.00){\oval(3.00,3.00)[rt]}
\put(193.50,81.00){\oval(3.00,3.00)[lb]}
\
\put(192.00,81.00){\line(0,-1){0.03}}
\put(216.00,81.00){\line(0,-1){0.05}}
\put(216.00,81.00){\line(0,-1){0.05}}
\put(192.00,81.00){\line(0,-1){0.05}}
\put(131.33,9.67){\oval(2.00,2.00)[r]}
\put(131.33,11.67){\oval(2.00,2.00)[l]}
\put(131.33,13.67){\oval(2.00,2.00)[r]}
\put(131.33,15.67){\oval(2.00,2.00)[l]}
\put(131.33,17.67){\oval(2.00,2.00)[r]}
\put(131.33,19.67){\oval(2.00,2.00)[l]}
\put(131.33,21.67){\oval(2.00,2.00)[r]}
\put(131.33,23.67){\oval(2.00,2.00)[l]}
\put(131.33,25.67){\oval(2.00,2.00)[r]}
\put(131.33,27.67){\oval(2.00,2.00)[l]}
\put(131.33,29.67){\oval(2.00,2.00)[r]}
\put(131.33,31.67){\oval(2.00,2.00)[l]}
\put(155.66,9.67){\oval(2.00,2.00)[r]}
\put(155.66,11.67){\oval(2.00,2.00)[l]}
\put(155.66,13.67){\oval(2.00,2.00)[r]}
\put(155.66,15.67){\oval(2.00,2.00)[l]}
\put(155.66,17.67){\oval(2.00,2.00)[r]}
\put(155.66,19.67){\oval(2.00,2.00)[l]}
\put(155.66,21.67){\oval(2.00,2.00)[r]}
\put(155.66,23.67){\oval(2.00,2.00)[l]}
\put(155.66,25.67){\oval(2.00,2.00)[r]}
\put(155.66,27.67){\oval(2.00,2.00)[l]}
\put(155.66,29.67){\oval(2.00,2.00)[r]}
\put(155.66,31.67){\oval(2.00,2.00)[l]}
\put(120.66,32.67){\line(1,0){45.67}}
\put(120.66,8.67){\line(1,0){45.67}}
\put(125.33,32.67){\vector(1,0){1.00}}
\put(136.67,32.67){\vector(1,0){1.00}}
\put(147.67,32.67){\vector(1,0){1.00}}
\put(160.33,32.67){\vector(1,0){1.00}}
\put(127.00,8.67){\vector(-1,0){1.33}}
\put(162.67,8.67){\vector(-1,0){1.33}}
\put(143.33,9.67){\oval(2.00,2.00)[r]}
\put(143.33,11.67){\oval(2.00,2.00)[l]}
\put(143.33,13.67){\oval(2.00,2.00)[r]}
\put(143.33,15.67){\oval(2.00,2.00)[l]}
\put(143.33,17.67){\oval(2.00,2.00)[r]}
\put(143.33,19.67){\oval(2.00,2.00)[l]}
\put(143.33,21.67){\oval(2.00,2.00)[r]}
\put(143.33,23.67){\oval(2.00,2.00)[l]}
\put(143.33,25.67){\oval(2.00,2.00)[r]}
\put(143.33,27.67){\oval(2.00,2.00)[l]}
\put(143.33,29.67){\oval(2.00,2.00)[r]}
\put(143.33,31.67){\oval(2.00,2.00)[l]}
\put(25.33,9.67){\oval(2.00,2.00)[r]}
\put(25.33,11.67){\oval(2.00,2.00)[l]}
\put(25.33,13.67){\oval(2.00,2.00)[r]}
\put(25.33,15.67){\oval(2.00,2.00)[l]}
\put(25.33,17.67){\oval(2.00,2.00)[r]}
\put(25.33,19.67){\oval(2.00,2.00)[l]}
\put(25.33,21.67){\oval(2.00,2.00)[r]}
\put(25.33,23.67){\oval(2.00,2.00)[l]}
\put(25.33,25.67){\oval(2.00,2.00)[r]}
\put(25.33,27.67){\oval(2.00,2.00)[l]}
\put(25.33,29.67){\oval(2.00,2.00)[r]}
\put(25.33,31.67){\oval(2.00,2.00)[l]}
\put(42.66,9.67){\oval(2.00,2.00)[r]}
\put(42.66,11.67){\oval(2.00,2.00)[l]}
\put(42.66,13.67){\oval(2.00,2.00)[r]}
\put(42.66,15.67){\oval(2.00,2.00)[l]}
\put(42.66,17.67){\oval(2.00,2.00)[r]}
\put(42.66,19.67){\oval(2.00,2.00)[l]}
\put(42.66,21.67){\oval(2.00,2.00)[r]}
\put(42.66,23.67){\oval(2.00,2.00)[l]}
\put(42.66,25.67){\oval(2.00,2.00)[r]}
\put(42.66,27.67){\oval(2.00,2.00)[l]}
\put(42.66,29.67){\oval(2.00,2.00)[r]}
\put(42.66,31.67){\oval(2.00,2.00)[l]}
\put(10.66,32.67){\line(1,0){45.67}}
\put(10.66,8.67){\line(1,0){45.67}}
\put(15.33,32.67){\vector(1,0){1.00}}
\put(50.33,32.67){\vector(1,0){1.00}}
\put(17.00,8.67){\vector(-1,0){1.33}}
\put(35.00,8.67){\vector(-1,0){1.33}}
\put(52.67,8.67){\vector(-1,0){1.33}}
\put(166.00,57.00){\line(-1,0){12.94}}
\put(166.00,81.00){\line(-1,0){12.94}}
\put(135.00,57.00){\line(-1,0){12.94}}
\put(135.00,81.00){\line(-1,0){12.94}}
\put(79.00,33.00){\line(0,-1){0.08}}
\put(97.00,33.00){\line(0,-1){0.08}}
\put(80.00,33.00){\oval(2.00,2.00)[t]}
\put(82.00,33.00){\oval(2.00,2.00)[b]}
\put(84.00,33.00){\oval(2.00,2.00)[t]}
\put(86.00,33.00){\oval(2.00,2.00)[b]}
\put(88.00,33.00){\oval(2.00,2.00)[t]}
\put(90.00,33.00){\oval(2.00,2.00)[b]}
\put(92.00,33.00){\oval(2.00,2.00)[t]}
\put(94.00,33.00){\oval(2.00,2.00)[b]}
\put(96.00,33.00){\oval(2.00,2.00)[t]}
\put(80.00,9.00){\oval(2.00,2.00)[t]}
\put(82.00,9.00){\oval(2.00,2.00)[b]}
\put(84.00,9.00){\oval(2.00,2.00)[t]}
\put(86.00,9.00){\oval(2.00,2.00)[b]}
\put(88.00,9.00){\oval(2.00,2.00)[t]}
\put(90.00,9.00){\oval(2.00,2.00)[b]}
\put(92.00,9.00){\oval(2.00,2.00)[t]}
\put(94.00,9.00){\oval(2.00,2.00)[b]}
\put(96.00,9.00){\oval(2.00,2.00)[t]}
\put(79.00,33.00){\line(0,-1){0.01}}
\put(97.00,33.00){\line(0,-1){0.01}}
\put(79.00,8.00){\line(0,0){0.00}}
\put(79.00,8.00){\line(0,0){0.00}}
\put(110.00,9.00){\line(-1,0){12.94}}
\put(110.00,33.00){\line(-1,0){12.94}}
\put(79.00,9.00){\line(-1,0){12.94}}
\put(79.00,33.00){\line(-1,0){12.94}}
\put(229.00,81.00){\line(-1,0){12.94}}
\put(229.00,57.00){\line(-1,0){12.94}}
\put(192.00,57.00){\line(-1,0){12.94}}
\put(192.00,81.00){\line(-1,0){12.94}}
\put(56.00,67.00){\makebox(0,0)[cc]{$(m)$}}
\put(113.00,67.00){\makebox(0,0)[cc]{$(n)$}}
\put(166.00,67.00){\makebox(0,0)[cc]{$(p)$}}
\put(230.00,66.00){\makebox(0,0)[cc]{$(q)$}}
\put(166.00,21.00){\makebox(0,0)[cc]{$(t)$}}
\put(111.00,22.00){\makebox(0,0)[cc]{$(s)$}}
\put(56.00,21.00){\makebox(0,0)[cc]{$(r)$}}
\put(135.00,81.00){\line(0,-1){0.10}}
\put(153.00,81.00){\line(0,-1){0.10}}
\put(73.66,9.67){\oval(2.00,2.00)[r]}
\put(73.66,11.67){\oval(2.00,2.00)[l]}
\put(73.66,13.67){\oval(2.00,2.00)[r]}
\put(73.66,15.67){\oval(2.00,2.00)[l]}
\put(73.66,17.67){\oval(2.00,2.00)[r]}
\put(73.66,19.67){\oval(2.00,2.00)[l]}
\put(73.66,21.67){\oval(2.00,2.00)[r]}
\put(73.66,23.67){\oval(2.00,2.00)[l]}
\put(73.66,25.67){\oval(2.00,2.00)[r]}
\put(73.66,27.67){\oval(2.00,2.00)[l]}
\put(73.66,29.67){\oval(2.00,2.00)[r]}
\put(73.66,31.67){\oval(2.00,2.00)[l]}
\put(79.00,33.00){\line(0,-1){0.01}}
\put(97.00,33.00){\line(0,-1){0.01}}
\put(79.00,33.00){\line(0,-1){24.00}}
\put(97.00,33.00){\line(0,-1){24.00}}
\put(135.00,81.00){\line(0,-1){24.00}}
\put(153.00,81.00){\line(0,-1){24.00}}
\put(192.00,81.00){\line(0,-1){24.00}}
\put(216.00,81.00){\line(0,-1){24.00}}
\put(69.00,33.00){\vector(1,0){1.00}}
\put(76.00,33.00){\vector(1,0){0.99}}
\put(79.00,23.00){\vector(0,-1){0.04}}
\put(77.00,9.00){\vector(-1,0){0.99}}
\put(70.00,9.00){\vector(-1,0){1.01}}
\put(97.00,21.00){\vector(0,1){0.00}}
\put(102.00,33.00){\vector(1,0){0.99}}
\put(104.00,9.00){\vector(-1,0){1.01}}
\put(151.00,8.50){\vector(-1,0){1.04}}
\put(138.00,8.50){\vector(-1,0){0.98}}
\put(127.00,81.00){\vector(1,0){1.05}}
\put(135.00,71.00){\vector(0,-1){0.00}}
\put(129.00,57.00){\vector(-1,0){0.95}}
\put(153.00,69.00){\vector(0,1){0.05}}
\put(159.00,81.00){\vector(1,0){1.02}}
\put(161.00,57.00){\vector(-1,0){0.98}}
\put(184.00,81.00){\vector(1,0){0.98}}
\put(192.00,71.00){\vector(0,-1){0.06}}
\put(185.00,57.00){\vector(-1,0){0.94}}
\put(216.00,69.00){\vector(0,1){0.02}}
\put(223.00,57.00){\vector(-1,0){0.98}}
\put(222.00,81.00){\vector(1,0){0.94}}
\put(33.00,80.50){\vector(1,0){1.00}}
\put(33.00,32.50){\vector(1,0){1.05}}
\put(89.00,80.50){\vector(1,0){0.95}}
\
\put(21.00,33.00){\oval(4.00,2.00)[lt]}
\put(21.00,35.00){\oval(4.00,2.00)[rb]}
\put(25.00,35.00){\oval(4.00,2.00)[lt]}
\put(25.00,37.00){\oval(4.00,2.00)[rb]}
\put(29.00,37.00){\oval(4.00,2.00)[lt]}
\put(29.00,39.00){\oval(4.00,2.00)[rb]}
\put(33.00,39.00){\oval(4.00,2.00)[lt]}
\
\put(45.00,33.00){\oval(4.00,2.00)[rt]}
\put(45.00,35.00){\oval(4.00,2.00)[lb]}
\put(41.00,35.00){\oval(4.00,2.00)[rt]}
\put(41.00,37.00){\oval(4.00,2.00)[lb]}
\put(37.00,37.00){\oval(4.00,2.00)[rt]}
\put(37.00,39.00){\oval(4.00,2.00)[lb]}
\put(33.00,39.00){\oval(4.00,2.00)[rt]}
\end{picture}
\caption{Box diagrams up to 2--loop level.}
\end{figure}

The second order PT contribution to the matrix element squared reads
\begin{eqnarray}
\Delta|M|^2 &=& 2(M_a+M_b)^*(M_{\mathrm{2-vertex}}) \nonumber \\
&+& |M_c+M_d+M_e+M_f+M_m+M_n+M_p+M_q|^2,
\end{eqnarray}
where $M_{\mathrm{2-vertex}}$ is the matrix element of
the ten 2-loop vertex Feynman diagrams.
The matrix element of elastic Bhabha scattering, including
relevant contributions up to the 2--loop level, may be written in the
following form:
\begin{eqnarray}
M&=&M_{0t}(1+\delta_t^{(1)}+\delta_t^{(2)})
- M_{0s}(1+\delta_s^{(1)}+\delta_s^{(2)})+B_1^{(1)}+B_2^{(1)}
- B_3^{(1)} - B_4^{(1)} + B^{(2)},
\end{eqnarray}
where
\begin{eqnarray*}
M_{0t}&=&\frac{4\pi\alpha \mathrm{i}}{t}\bar u(p_1')\gamma_{\mu}
u(p_1)\bar v(p_2)\gamma_{\mu}v(p_2'), \quad
M_{0s}=\frac{4\pi\alpha \mathrm{i}}{t}\bar v(p_2')\gamma_{\mu}
u(p_1)\bar u(p_1')\gamma_{\mu}v(p_2).
\end{eqnarray*}
The quantities $B_{i}^{(1)}$ correspond to the 1--loop box--type diagrams
(see Fig.~2(m--q)), whereas $B^{(2)}$ comes from the 2--loop ones
(some representatives are drawn in Fig.~2(r--t)).
At the Born level we have
\begin{eqnarray}
\sum|M_{0t}^2|&=&(4\pi\alpha)^2\frac{8}{t^2}(s^2+u^2),\quad
\sum|M_{0s}^2|=(4\pi\alpha)^2\frac{8}{s^2}(t^2+u^2),\quad \nonumber \\
\sum M_{0s}M_{0t}^*&=&-(4\pi\alpha)^2 8\frac{u^2}{st},\quad
\sum|M_{0t}-M_{0s}|^2=16(4\pi\alpha)^2\left(\frac{s}{t}+\frac{t}{s}
+1\right)^2\,, \\
&&s=(p_1+p_2)^2,\qquad t=(p_1-p_1')^2,\qquad u=(p_1-p_2')^2,\nonumber \\
&&p_1+p_2=p_1'+p_2',\qquad p_{1,2}^2=p_{1,2}'^2=m^2. \nonumber
\end{eqnarray}
The quantities $\delta_t^{(1)}, \delta_t^{(2)}$ are real. They read
\begin{eqnarray}\label{GP}
\delta_t^{(1)}&=&\frac{\alpha}{\pi}\left[2\Gamma^{(2)}(t)
+\Pi^{(2)}(t)\right],\nonumber \\
\delta_t^{(2)}&=&\left(\frac{\alpha}{\pi}\right)^2[(\Gamma^{(2)}(t))^2
+2\Pi^{(2)}(t)\Gamma^{(2)}(t)+(\Pi^{(2)}(t))^2+\Pi^{(4)}(t)+2\Gamma^{(4)}(t)].
\end{eqnarray}
Here $\Pi^(i)(t)$ are the vacuum polarization insertions.
Similar expressions are held for $\delta_s^{(1)}, \delta_s^{(2)}$ and
can be derived from~(\ref{GP}) by using crossing relations (relevant
quantities are, in general, complex valued).

The relevant second order PT contribution to the matrix element, squared
and summed over spin states, looks as follows:
\begin{equation}
\Delta\sum|M|^2=\alpha^2\left(\frac{\alpha}{\pi}\right)^2
(\Delta_1+\Delta_2+\Delta_3+\Delta_4),
\end{equation}
where
\begin{eqnarray} \label{deldef}
\frac{\alpha^4}{\pi^2}\Delta_1&=&\sum |M_{0t}|^2(|\delta_t^{(1)}|^2
+ 2\delta_t^{(2)}) + \sum |M_{0s}|^2((\delta_s^{(1)})^2
+ 2\mathrm{Re}\; \delta_s^{(2)}) \nonumber \\
&-& 2\mathrm{Re}\; \sum M_{0s}^*M_{0t}(\delta_t^{(1)}\delta_s^{(1)}
+\delta_s^{(2)}+\delta_t^{(2)}), \nonumber \\
\frac{\alpha^4}{\pi^2}\Delta_2&=&2\mathrm{Re}\; \sum(M_{0t}\delta_t^{(1)}-M_{0t}\delta_t^{(1)})^*
(B_1^{(1)}+B_2^{(1)}-B_3^{(1)}-B_4^{(1)}),  \nonumber \\
\frac{\alpha^4}{\pi^2}\Delta_3&=&\sum|B_1^{(1)}+B_2^{(1)}-B_3^{(1)}-B_4^{(1)}|^2, \nonumber \\
\frac{\alpha^4}{\pi^2}\Delta_4&=&2\mathrm{Re}\; \sum(M_{0t}-M_{0s})^*B^{(2)}.
\end{eqnarray}
Quantity $B^{(2)}$, which enters into the definition of $\Delta_4$,
is nowadays unknown. The aim of this work is to give explicit expressions
for $\Delta_1$, $\Delta_2$, and  $\Delta_3$. The first term $\Delta_1$
was given above. As for $\Delta_2$, using the usual notation
$\hat a = \gamma_{\mu}a^{\mu}$, it can be cast
down in the form
\begin{eqnarray} \label{delta2}
\Delta_2&=&(1+{\cal{P}}_{st})\mathrm{Re}\; \frac{\Gamma^{(2)}(t)}{t}
\int\frac{\dd^4k}{\mathrm{i}\pi^2}\biggl\{
\frac{{\mathrm{Tr}}\;(\gamma_{\mu}(\hat p_1+\hat k)\gamma_{\nu}\hat p_1
\gamma_{\rho}\hat p_1')\;\;
{\mathrm{Tr}}\;(\gamma_{\mu}(-\hat p_2+\hat k)\gamma_{\nu}\hat p_2
\gamma_{\rho}\hat p_2')}{A(p_1+k)A(k)A(k-q)A(-p_2+k)} \nonumber \\
&+&\frac{{\mathrm{Tr}}\;(\gamma_{\mu}(\hat p_1+\hat k)\gamma_{\nu}
\hat p_1\gamma_{\rho}\hat p_1')\;\;
{\mathrm{Tr}}\;(\gamma_{\mu}(-\hat p_2'-\hat k)\gamma_{\nu}\hat p_2'
\gamma_{\rho}\hat p_2)}{A(p_1+k)A(k)A(k-q_1)A(p_2'+k)} \nonumber \\
&+& \frac{2{\mathrm{Tr}}\;(\hat p_2\hat p_1'(\hat p_1+\hat k)\hat p_2'\hat p_1
(\hat p_1'+\hat k))}{A(p_1+k)A(k)A(k-q_1)A(p_1'+k)}
+ \frac{2u^2(p_1+k)(-p_2'-k)}{A(p_1+k)A(k)A(k-q_1)A(p_2'+k)} \biggr\}
+ \delta\Delta_2, \nonumber \\
&& A(p_{1,2}\pm k)=(p_{1,2}\pm k)^2-m^2, \quad
A(p_{1,2}'\pm k)=(p_{1,2}'\pm k)^2-m^2, \quad q=p_1'-p_1, \nonumber \\
&& A(q-k)=(q-k)^2-\lambda^2, \quad A(q_1-k)=(q_1-k)^2-\lambda^2,
\quad q_1=-p_1-p_2\, . \nonumber
\end{eqnarray}
where the permutation operator ${\cal{P}}_{st}$ acts as follows:
$$
{\cal{P}}_{st}A(s,t,u,L,L_t)=A(t,s,u,L_t,L)
$$
Calculating the first term in $\Delta_2$, we have to put (see Appendix A)
\begin{equation} \label{rules}
L_s=L,\qquad\psi_{1s}=\frac{1}{s}\left(\frac{1}{2}L^2+\frac{\pi^2}{6}\right).
\end{equation}
The second term on the right hand side (rhs) of Eq.~(\ref{delta2})
$\delta\Delta_2$
arises from the product of imaginary parts of $(\Gamma^{(2)}(s))^*$
and box structures (see Eq.~(\ref{deldef}) and Appenix~A).
It can be obtained by applying the following rules:
\begin{eqnarray} \label{rules1}
&&\delta {\mathrm{Re}}\;  \Gamma^{(2)}(s)^*L_s=-\pi^2\left(\frac{1}{2}L
-\frac{1}{2}L_{\lambda}-\frac{3}{4}\right),
\quad \delta {\mathrm{Re}}\; \Gamma^{(2)}(s)^*\psi_{1s}=-\frac{\pi^2}{s}L
\left(\frac{1}{2}L-\frac{1}{2}L_{\lambda}-\frac{3}{4}\right), \nonumber \\
&&\delta {\mathrm{Re}}\; L_s^*L_s=\pi^2, \quad
\delta {\mathrm{Re}}\; L_s^*\psi_{1s}=\frac{\pi^2L}{s}\, , \quad
\delta {\mathrm{Re}}\; \psi_{1s}^*\psi_{1s}= \frac{\pi^2L^2}{s^2}\, .
\end{eqnarray}
By performing the loop--momentum integration (relevant integrals are
given in Appendix~A) one arrives to the result (see Eq.~(\ref{deltai})).

Consider now $\Delta_3$. The symmetry properties
permit us to express it in the form
\begin{eqnarray}
\Delta_3&=&|B_1+B_2-B_3-B_4|^2=[1+{\cal{P}}_{su}+
(1+{\cal{P}}_{tu}){\cal{P}}_{st}]|B_1|^2+2(1+{\cal{P}}_{st})[B_1B_2^*-
B_2B_3^* \nonumber \\
&-& \frac{1}{2}B_1B_3^*-\frac{1}{2}B_2B_4^*]+\delta\Delta_3, \\ \nonumber
&& {\cal{P}}_{su}A(s,t,u,L,L_t)=A(u,t,s,L_u,L_t), \qquad
{\cal{P}}_{tu}A(s,t,u,L,L_t)=A(s,u,t,L,L_u).
\end{eqnarray}
The quantity $\delta\Delta_3$ is to be written according to the
rules mentioned earlier~(\ref{rules1}). In calculations of the first
two terms in $\Delta_3$ we have to take $L_s$ and $\psi_{1s}$ as
in~(\ref{rules}).
The remaining contributions to $\Delta_3$ are
\begin{eqnarray}
|B_1|^2 &=& \int\frac{\dd^4k_1}{\mathrm{i}\pi^2}
\int\frac{\dd^4k}{\mathrm{i}\pi^2}
\frac{{\mathrm{Tr}}\;(\hat p_1'\gamma_{\mu}(\hat p_1+\hat k)\gamma_{\nu}
\hat p_1\gamma_{\xi}(\hat p_1+\hat k_1)
\gamma_{\eta})}{A(p_1+k_1)A(k_1)A(k_1-q)A(-p_2+k_1)} \nonumber \\
&\times& \frac{{\mathrm{Tr}}\;(\hat p_2\gamma_{\nu}(-\hat p_2-\hat k)
\gamma_{\mu}\hat p_2'\gamma_{\xi}(-\hat p_2'+\hat k_1)\gamma_{\eta})}
{A(p_1+k)A(k)A(k-q)A(p_2-k)}\, , \nonumber \\
B_1B_2^*&=&\int\frac{\dd^4k_1}{\mathrm{i}\pi^2}
\int\frac{\dd^4k}{\mathrm{i}\pi^2}
\frac{{\mathrm{Tr}}\;(\hat p_1'\gamma_{\mu}(\hat p_1+\hat k)\gamma_{\nu}
\hat p_1\gamma_{\xi}(\hat p_1+\hat k_1)\gamma_{\eta})}
{A(p_1+k)A(k)A(k-q)A(p_2-k)}  \nonumber \\
&\times& \frac{{\mathrm{Tr}}\;(\hat p_2\gamma_{\nu}(-\hat p_2+\hat k)
\gamma_{\mu}\hat p_2'\gamma_{\xi}(-\hat p_2'-\hat k_1)\gamma_{\eta})}
{A(p_1+k_1)A(k_1)A(k_1-q)A(p_2'+k_1)}\, , \\
B_2B_3^*&=&\int\frac{\dd^4k_1}{\mathrm{i}\pi^2}
\int\frac{\dd^4k}{\mathrm{i}\pi^2} \nonumber \\
&\times& \frac{{\mathrm{Tr}\; }\bigl(\hat p_1'\gamma_{\mu}
(\hat p_1+\hat k)\gamma_{\nu}\hat p_1\gamma_{\eta}(\hat p_1+\hat k_1)
\gamma_{\xi}\hat p_2\gamma_{\mu}(-\hat p_2-\hat k)
\gamma_{\nu}\hat p_2'\gamma_{\xi}(\hat p_1'+\hat k_1)\gamma_{\eta}\bigr)}
{A(p_1+k)A(k)A(k-q)A(p_2'+k)A(p_1+k_1)A(k_1)A(k_1-q_1)A(p_1'+k_1)}\, ,
\nonumber \\
B_1B_3^*&=&\int\frac{\dd^4k_1}{\mathrm{i}\pi^2}
\int\frac{\dd^4k}{\mathrm{i}\pi^2}  \nonumber \\
&\times& \frac{{\mathrm{Tr}\; }\bigl(\hat p_1'\gamma_{\mu}(\hat p_1+\hat k)
\gamma_{\nu}\hat p_1\gamma_{\eta}(\hat p_1+\hat k_1)
\gamma_{\xi}\hat p_2\gamma_{\nu}(-\hat p_2+\hat k)
\gamma_{\mu}\hat p_2'\gamma_{\xi}(\hat p_1'+\hat k_1)\gamma_{\eta}\bigr)}
{A(p_1+k)A(k)A(k-q)A(p_2-k)A(p_1+k_1)A(k_1)A(k_1-q_1)A(p_1'+k_1)}\, ,
\nonumber \\
B_2B_4^*&=&\int\frac{\dd^4k_1}{\mathrm{i}\pi^2}
\int\frac{\dd^4k}{\mathrm{i}\pi^2} \nonumber \\
&\times& \frac{{\mathrm{Tr}\;}\bigl(\hat p_1'\gamma_{\mu}(\hat p_1+\hat k)\gamma_{\nu}\hat p_1\gamma_{\eta}(\hat p_1+\hat k_1)
\gamma_{\xi}\hat p_2\gamma_{\mu}(-\hat p_2'-\hat k)\gamma_{\nu}\hat p_2'
\gamma_{\eta}(-\hat p_2'-\hat k_1)\gamma_{\xi}\bigr)}
{A(p_1+k)A(k)A(k-q)A(p_2'+k)A(p_1+k_1)A(k_1)A(k_1-q_1)A(p_2'+k_1)}\, .
\nonumber
\end{eqnarray}
Standard but rather tedious computation gives the following result:
\begin{eqnarray} \label{deltai}
\Delta_i&=&L_{\lambda}^2(a_{i1}L^2+a_{i2}L)
+L_{\lambda}(a_{i3}L^3+a_{i4}L^2+a_{i5}L)+a_{i6}L^4 \nonumber \\
&+& a_{i7}L^3+a_{i8}L^2+a_{i9}L, \qquad i=1,2,3.
\end{eqnarray}
Coefficients $a_{ij}$ as the functions of
$\theta=\widehat{\vecc{p}_1,\vecc{p}}_1'$ are given below:
\begin{eqnarray*}
a_{11}&=&2F^2,\qquad a_{12}=-4F^2-2\left[F+2\left(
\frac{s^2}{t^2}+\frac{s}{t}+1\right)\right]L_{st}, \nonumber \\
a_{13}&=&-2F^2,\qquad a_{14}=\frac{28}{3}F^2+\left[3F
+6\left(\frac{s^2}{t^2}+\frac{s}{t}+1\right)\right]L_{st}, \nonumber \\
a_{15}&=&\left(-6\frac{s^2}{t^2}-8\frac{s}{t}-2\frac{t}{s}-7\right)L_{st}^2
+\left(-\frac{56}{3}\frac{s^2}{t^2}-\frac{28}{3}\frac{t}{s}-28\frac{s}{t}
-28\right)L_{st} \nonumber \\
&+& \left(\frac{2}{3}\frac{s^2}{t^2}+\frac{10}{3}
\frac{s}{t}+\frac{28}{3}\frac{t}{s}+\frac{20}{3}\frac{t^2}{s^2}+9\right)\pi^2
-\frac{158}{9}F^2, \nonumber \\
a_{16}&=&\frac{1}{2}F^2,\quad a_{17}=-\frac{11}{3}F^2
-\left(2\frac{s^2}{t^2}+3\frac{s}{t}+\frac{t}{s}+3\right)L_{st},\nonumber \\
a_{18}&=&\left(3\frac{s^2}{t^2}+4\frac{s}{t}+\frac{t}{s}
+\frac{7}{2}\right)L_{st}^2+\left[\frac{11}{2}F+11\left(\frac{s^2}{t^2}
+\frac{s}{t}+1\right)\right]L_{st} \nonumber \\
&-&\left(\frac{1}{3}\frac{s^2}{t^2}
+\frac{5}{3}\frac{s}{t}+\frac{14}{3}\frac{t}{s}+\frac{10}{3}\frac{t^2}{s^2}
+\frac{9}{2}\right)\pi^2+\frac{215}{18}F^2,\nonumber \\
a_{19}&=&\left(-2\frac{s^2}{t^2}-\frac{5}{2}\frac{s}{t}-\frac{1}{2}
\frac{t}{s}-2\right)L_{st}^3-\left(11\frac{s^2}{t^2}+\frac{44}{3}\frac{s}{t}
+\frac{11}{3}\frac{t}{s}+\frac{77}{6}\right)L_{st}^2 \nonumber \\
&+&\left(
\frac{2}{3}\frac{s^2}{t^2}+\frac{3}{2}\frac{s}{t}+\frac{5}{6}\frac{t}{s}
+2\right)\pi^2L_{st}-\frac{215}{3}\left(\frac{1}{3}\frac{s^2}{t^2}
+\frac{1}{2}\frac{s}{t}+\frac{1}{6}\frac{t}{s}+\frac{1}{2}\right)L_{st} \nonumber \\
&+&\left(\frac{13}{18}\frac{s^2}{t^2}+\frac{85}{18}\frac{s}{t}+
\frac{257}{18}\frac{t}{s}+\frac{185}{18}\frac{t^2}{s^2}+\frac{27}{2}
\right)\pi^2-\left(\frac{1313}{72}-6\zeta(3)\right)F^2,\nonumber
\end{eqnarray*}
\begin{eqnarray*}
a_{21}&=& 0, \qquad a_{22}=4F^2L_{su}-2\left[F+2\left(\frac{t^2}{s^2}
+\frac{t}{s}+1\right)\right]L_{st}, \nonumber \\
a_{23}&=& 0, \quad a_{24}=-6F^2L_{su}+3\left[F+2\left(\frac{t^2}{s^2}
+\frac{t}{s}+1\right)\right]L_{st}, \nonumber \\
a_{25} &=& \left(12\frac{s^2}{t^2}+20\frac{s}{t}+6\frac{t}{s}+19\right)L_{st}L_{su}
+\left(\frac{34}{3}F^2-\frac{s}{t}-\frac{t}{s}\right)L_{su} \nonumber \\
&-&\frac{1}{3}\left(34F^2-17F-34\frac{s^2}{t^2}-37\frac{s}{t}-34\right)L_{st}
-\left(\frac{11}{2}F-\frac{s}{t}-6\right)L_{st}^2-FL_{su}^2 \nonumber \\
&+&\pi^2\left(4F-\frac{s}{t}+\frac{7}{2}\right), \qquad
a_{26}=0, \quad a_{27}=2F^2L_{su}-\left[F+2\left(\frac{t^2}{s^2}
+\frac{t}{s}+1\right)\right]L_{st}, \nonumber \\
a_{28}&=&\left[-2F\left(3\frac{s}{t}+2\right)+4\frac{t}{s}+\frac{1}{2}\right]
L_{st}L_{su}+\left[\frac{11}{4}F+\frac{1}{2}\frac{s}{t}+3\right]L_{st}^2
+\frac{1}{2}FL_{su}^2-\biggl[\frac{22}{3}F^2  \nonumber \\
&-&\frac{1}{2}\left(\frac{s}{t}+\frac{t}{s}\right)\biggr]L_{su}
+\left[\frac{19}{6}F+\frac{22}{3}\left(\frac{t^2}{s^2}+\frac{t}{s}+1\right)
+\frac{1}{2}\frac{t}{s}+\frac{1}{2}\right]L_{st}
-\pi^2\left[2F+\frac{1}{2}\frac{s}{t}-\frac{7}{4}\right], \nonumber \\
a_{29}&=&-\left(F+\frac{5}{2}\frac{s}{t}+3\right)L_{st}^3
+\left(6\frac{s^2}{t^2}+9\frac{s}{t}+\frac{t}{s}+6\right)L_{st}^2L_{su}
-\left(\frac{s}{t}+\frac{1}{2}\right)L_{st}L_{su}^2 \nonumber \\
&+&\frac{1}{3}\left(44\frac{s^2}{t^2}+74\frac{s}{t}+22\frac{t}{s}+\frac{143}{2}
\right)L_{st}L_{su}
-\left(\frac{77}{12}F-\frac{5}{6}\frac{s}{t}-\frac{22}{3}
\right)L_{st}^2 \nonumber \\
&-&\frac{11}{6}FL_{su}^2-\left(\frac{2}{3}F+
\frac{18}{3}\frac{t^2}{s^2}+\frac{s}{t}+9\frac{t}{s}+6\right)\pi^2L_{su} \nonumber \\
&+&\left[\frac{92}{9}F^2-\frac{11}{6}\left(\frac{s}{t}+\frac{t}{s}\right)
\right]L_{su}+\left(\frac{20}{3}\frac{t^2}{s^2}+6\frac{t}{s}
+\frac{4}{3}\frac{s}{t}+\frac{9}{2}\right)\pi^2L_{st}
-\biggl(\frac{46}{3}F+\frac{92}{9}\frac{t^2}{s^2} \nonumber \\
&-&\frac{217}{18}\frac{s}{t}\biggr)L_{st}
+\left(\frac{11}{6}\frac{s}{t}
+\frac{8}{3}\frac{t}{s}+\frac{77}{12}\right)\pi^2, \nonumber
\end{eqnarray*}
\begin{eqnarray*}
a_{31}&=&0,\quad a_{32}=0,\quad a_{33}=0, \quad a_{34}=0, \quad
a_{35}=-2F^2L_{su}^2+2F\left(2\frac{t}{s}+1\right)L_{st}L_{su} \nonumber \\
&-&\left(2\frac{t^2}{s^2}+2\frac{t}{s}+1\right)L_{st}^2
-\left(2\frac{s^2}{t^2}+2\frac{s}{t}+1\right)\pi^2,
\quad a_{36}=0,\quad a_{37}=0, \nonumber \\
a_{38}&=&F^2L_{su}^2-F\left(2\frac{t}{s}+1
\right)L_{st}L_{su}+\left(\frac{t^2}{s^2}+\frac{t}{s}+\frac{1}{2}\right)
L_{st}^2
+\pi^2\left(\frac{s^2}{t^2}+\frac{s}{t}+\frac{1}{2}\right), \nonumber \\
a_{39}&=&\frac{1}{2}FL_{su}^3+\frac{1}{2}\left(\frac{s}{t}+\frac{t}{s}\right)
(L_{su}^2-L_{st}L_{su})-\biggl[F\left(2\frac{s}{t}+\frac{3}{2}\right)
+\frac{1}{2}\left(\frac{s}{t}+\frac{1}{2}\right)\biggr]L_{su}^2L_{st}\nonumber \\
&+&\left[\frac{7}{4}F+\frac{1}{2}\left(\frac{s}{t}+4\right)\right]L_{st}^2L_{su}
+\left(F+\frac{1}{2}\frac{s}{t}+\frac{5}{4}\right)\pi^2L_{su}
-\frac{3}{4}\left(\frac{t}{s}+1\right)L_{st}^3 \nonumber \\
&-&\biggl[F^2+\left(\frac{s}{t}+\frac{t}{s}\right)\left(\frac{s}{t}
-\frac{t}{s}-3\right)
-\frac{1}{2}\left(\frac{s}{t}+9\right)\biggr]\pi^2L_{st}, \\
{\mathrm{where}} && \qquad F=\frac{s}{t}+\frac{t}{s}+1,\qquad
L_{st} = \ln\frac{s}{-t}\, , \qquad L_{su} = \ln\frac{s}{-u}\, .
 \nonumber
\end{eqnarray*}

\section{Emission of soft photons}

Second order corrections to the 1--loop virtual photon emission
corrected cross section, which arise from emission of
a single real soft photon having energy less than
$\Delta\varepsilon$, can be written down in the factorized form:
\begin{eqnarray}
\frac{\dd \sigma^{SV}}{\dd \sigma_0}&=&\frac{\alpha}{\pi}\delta_S
\frac{\alpha}{\pi}\delta_V=\left(\frac{\alpha}{\pi}\right)^2\Delta_{SV},
\qquad \dd\sigma_0=\frac{\alpha^2}{s}
\left(\frac{1-\chi+\chi^2}{\chi}\right)^2,
\nonumber \\ &&\chi=\frac{1}{2}(1-\cos\theta)=\sin^2\frac{\theta}{2},
\quad \theta=(\widehat{\vecc{p}_1,\vecc{p}}_1'), \qquad
\chi = \frac{-t}{s}\, \quad 1-\chi = \frac{-u}{s}\, ,
\nonumber \\
\delta_S&=&4\ln\frac{m\Delta\varepsilon}{\lambda\varepsilon}
\left(L-1+\ln\frac{\chi}{1-\chi}\right)
+L^2+2L\ln\frac{\chi}{1-\chi}+\ln^2\chi-\ln^2(1-\chi) \\ \nonumber
&-&\frac{2\pi^2}{3}+2\Li(1-\chi)-2\Li(\chi), \qquad
\Li(x) = - \int\limits_{0}^{x}\frac{\dd y}{y}\ln(1-y),
\\ \nonumber
\delta_V&=&4\ln\frac{m}{\lambda}\left(1-L+\ln\frac{1-\chi}{\chi}\right)
-L^2+2L\ln\frac{1-\chi}{\chi}-\ln^2\chi+\ln^2(1-\chi) \\ \nonumber
&+& 3L-4+f(\chi),\\
f(\chi)&=&(1-\chi+\chi^2)^{-2}\biggl[\frac{\pi^2}{12}(-4+8\chi
+3\chi^2-10\chi^3
+8\chi^4)+\frac{1}{2}(-2+5\chi-7\chi^2+5\chi^3  \nonumber \\
&-&2\chi^4)\ln^2(1-\chi)
+\frac{1}{4}\chi(3-\chi-3\chi^2+4\chi^3)\ln^2\chi
+\frac{1}{6}(22-30\chi+33\chi^2 \nonumber \\
&-&11\chi^3)\ln\chi-\frac{1}{2}\chi(1+\chi^2)\ln(1-\chi)
+\frac{1}{2}(4-8\chi+7\chi^2-2\chi^3)\ln\chi\ln(1-\chi) \biggr].\nonumber
\end{eqnarray}
The virtual corrections due to vacuum polarization were not taken
into account in the expression for $\delta_V$. They give an additional
contribution to the latter that looks:
\begin{equation}
\delta_{V\Pi}=\frac{2}{3}L-\frac{10}{9}-\frac{1}{3}(1-\chi+\chi^2)^{-2}
(2-3\chi+3\chi^2-\chi^3)\ln\chi.
\end{equation}
Consideration of emission of two soft photons having total energy
$\omega_1+\omega_2\leq\Delta\varepsilon$ requires some caution.
The final result (details of computations see in Appendix~B) has the
following form:
\begin{equation} \label{SS}
\frac{\dd \sigma^{SS}}{\dd \sigma_0}=\frac{1}{2!}\left(\frac{\alpha}{\pi}
\right)^2\biggl[\delta_S^2-\frac{8}{3}\pi^2\left(L-1+\ln\frac{\chi}{1-\chi}
\right)^2\biggr]\equiv \left(\frac{\alpha}{\pi}\right)^2\Delta_{SS}.
\end{equation}
Note that in the case photons are emitted independently, i.e.
$\omega_1\leq\Delta\varepsilon$ and $\omega_2\leq\Delta\varepsilon$,
the second term in square brackets will be absent. The multiplier $1/2!$
is due to the identity of photons.

\section{Conclusions}

At first we would like to mention that using a set of integrals
given in this paper one can perform
a similar calculation for the case of M\"oller scattering.

What one would expect from the real 2--box amplitudes contribuiton is
that the total correction must be free of infrared divergences
supplying by cancellation of fourth and third power of large logarithms.
We quote for completeness our final result derived by presenting it
in the form
\begin{eqnarray}
\Delta_{SV}+\Delta_{SS}&=&F^2\Biggl\{ L_{\lambda}^2L^2\cdot (-2)
+ L_{\lambda}^2L\cdot 4\left(1-\ln\frac{\chi}{1-\chi}\right) \nonumber \\
&+&L_{\lambda}L^3\cdot 2
+L_{\lambda}L^2\cdot \left(-8+6\ln\frac{\chi}{1-\chi}\right)
+L_{\lambda}L\cdot a_1 \nonumber \\
&-&L^4\cdot \frac{1}{2}
+L^3\cdot\left(3-2\ln\frac{\chi}{1-\chi}\right)
+L^2\cdot a_2+L\cdot b \\
&+&\ln\frac{\Delta\varepsilon}{\varepsilon}\cdot\left(12L^2+z_1L\right)
+\ln^2\frac{\Delta\varepsilon}{\varepsilon}\cdot\left(8L^2+z_2L\right)
 \Biggr\}, \nonumber \\
a_1&=&14+4\ln^2\left(\frac{\chi}{1-\chi}\right)
-10\ln\frac{\chi}{1-\chi}+2\ln^2\chi+2\ln^2(1-\chi)-2f(\chi), \nonumber \\
a_2&=&-4-\frac{8}{3}\pi^2-2\ln^2\left(\frac{\chi}{1-\chi}\right)
+6\ln\frac{\chi}{1-\chi}-\ln^2\chi+\ln^2(1-\chi)+f(\chi), \nonumber \\
b&=&-\frac{16}{3}\pi^2\ln\frac{\chi}{1-\chi}+\frac{10}{3}\pi^2
-2\ln\frac{\chi}{1-\chi}\left(\ln^2\chi-\ln^2(1-\chi)
-f(\chi)+4\right)  \nonumber \\
&+&3\ln^2\chi-3\ln^2(1-\chi)+6\Li(1-\chi)-6\Li(\chi),\nonumber \\
z_1&=&12\ln\frac{\chi}{1-\chi}-\frac{8}{3}\pi^2+8\Li(1-\chi)
-8\Li(\chi)+4f(\chi), \nonumber \\
z_2&=&-16\left(1-\ln\frac{\chi}{1-\chi}\right),
\nonumber \\ \nonumber \\
\sum\limits_{i=1}^3\Delta_i&=&F^2\biggl[L_{\lambda}^2L^2\cdot 2
+ L_{\lambda}^2L\cdot 4(L_{su}-L_{st}-1)
-L_{\lambda}L^3\cdot 2
+L_{\lambda}L^2\cdot \biggl(6L_{st}-6L_{su} \nonumber \\
&+& \frac{28}{3}\biggr) + L^4\cdot \frac{1}{2}
+L^3\cdot \left(2L_{su}-2L_{st}-\frac{11}{3}\right)\biggr]
+L_{\lambda}L\cdot c_1+L^2\cdot c_2+L\cdot d, \\
c_1&=&L_{st}^2\left(-6\frac{s^2}{t^2}-2\frac{t^2}{s^2}-\frac{25}{2}
\frac{s}{t}-\frac{19}{2}\frac{t}{s}-\frac{15}{2}\right)
+L_{st}L_{su}\biggl(12\frac{s^2}{t^2}+4\frac{t^2}{s^2}+22\frac{s}{t}
+12\frac{t}{s} \nonumber \\
&+&25\biggr)-L_{st}\left(\frac{56}{3}\frac{s^2}{t^2}+\frac{34}{3}
\frac{t^2}{s^2}+\frac{79}{3}\frac{t}{s}+\frac{98}{3}\frac{s}{t}+45\right)
+L_{su}\left(\frac{34}{3}F^2-F+1\right) \nonumber \\
&+&\pi^2\left(-\frac{4}{3}\frac{s^2}{t^2}+\frac{20}{3}
\frac{t^2}{s^2}+\frac{13}{3}\frac{s}{t}+\frac{40}{3}\frac{t}{s}
+\frac{31}{2}\right)-\frac{158}{9}F^2, \nonumber \\
c_2&=&L_{st}^2\left(3\frac{s^2}{t^2}+\frac{t^2}{s^2}+\frac{29}{4}
\frac{s}{t}+\frac{19}{4}\frac{t}{s}+\frac{39}{4}\right)
+\frac{3}{2}F^2L_{su}^2+L_{st}L_{su}\biggl(-2F^2 \nonumber \\
&-&F\left(4\frac{s}{t}-3\right)+4\frac{t}{s}+\frac{1}{2}\biggr)
+L_{st}\left(11F^2-\frac{7}{3}F-\frac{11}{3}
\left(\frac{t^2}{s^2}+\frac{t}{s}+1\right)+\frac{1}{2}\frac{t}{s}
+\frac{1}{2}\right)  \nonumber \\
&-&L_{su}\left(\frac{22}{3}F^2-\frac{1}{2}(F+1)\right)
-\pi^2\left(\frac{1}{3}\frac{s^2}{t^2}+\frac{4}{3}
\frac{t^2}{s^2}+\frac{25}{6}\frac{s}{t}+\frac{14}{3}\frac{t}{s}
+\frac{7}{4}\right)-\frac{215}{18}F^2, \nonumber \\
d&=&L_{st}^3\left(-2\frac{s^2}{t^2}-6\frac{s}{t}-
\frac{9}{4}\frac{t}{s}-\frac{27}{4}\right)+\frac{1}{2}F^2L_{su}^3
+L_{st}^2L_{su}\left(6\frac{s^2}{t^2}+\frac{45}{4}\frac{s}{t}
+\frac{11}{4}\frac{t}{s}+\frac{39}{4}\right) \nonumber \\
&-&L_{st}L_{su}^2\left(F\left(2\frac{s}{t}+\frac{3}{2}\right)
+\frac{3}{2}\frac{s}{t}+\frac{3}{4}\right)-L_{st}^2\left(
11\frac{s^2}{t^2}+\frac{81}{4}\frac{s}{t}+\frac{121}{12}\frac{t}{s}
+\frac{143}{12}\right) \nonumber \\
&-&L_{su}^2\left(\frac{4}{3}F+\frac{1}{2}\right)
+L_{st}L_{su}\left(\frac{44}{3}\frac{s^2}{t^2}
+\frac{145}{6}\frac{s}{t}+\frac{41}{6}\frac{t}{s}+\frac{143}{6}\right)
+\pi^2L_{st}\biggl(-\frac{4}{3}\frac{s^2}{t^2}\nonumber \\
&+&\frac{20}{3}\frac{t^2}{s^2}
+\frac{25}{3}\frac{s}{t}+\frac{71}{6}\frac{t}{s}+8\biggr)
+\pi^2L_{su}\left(-\frac{18}{3}\frac{s^2}{t^2}
-\frac{1}{6}\frac{s}{t}-\frac{26}{3}\frac{t}{s}-\frac{53}{12}\right)\nonumber \\
&-&L_{st}\left(\frac{46}{3}F+\frac{215}{9}\frac{s^2}{t^2}
+\frac{92}{9}\frac{t^2}{s^2}+\frac{215}{18}\frac{t}{s}+\frac{214}{9}
\frac{s}{t}-\frac{215}{6}\right) \nonumber \\
&+&L_{su}\left(\frac{92}{9}F^2
-\frac{11}{6}\frac{t}{s}-\frac{11}{6}\frac{s}{t}\right)
+\pi^2\left(\frac{13}{18}\frac{s^2}{t^2}+\frac{185}{18}\frac{t^2}{s^2}
+\frac{59}{9}\frac{s}{t}+\frac{305}{18}\frac{t}{s}+\frac{239}{12}\right)
\nonumber \\
&-&\left(\frac{1313}{72}-6\zeta(3)\right)F^2. \nonumber
\end{eqnarray}
Here $\Delta_{SS}$ and $\Delta_{SV}$ denote quasi--elastic
contributions, coming from double soft and soft--virtual photons emission.
It immediately follows that all the terms proportional to $L^4$,
$L_{\lambda}^2L^2$, $L_{\lambda}L^3$ and $L_{\lambda}^2L$ disappear
in the total sum. One must expect the cancellation
of the third power of large logarithms as well as the rest of infrared
singularities when contribution of 2--box diagrams will be taken
into account. As for the terms containing $L^2\ln^2(\Delta\varepsilon
/\varepsilon)$ and $L^2\ln(\Delta\varepsilon/\varepsilon)$,
they are explicitly seen to agree with those which could be
derived in the renormalization group approach. To show this,
let us write down the expression for cross section
according to the renormalization group:
\begin{eqnarray}
&&\frac{\dd \sigma}{\dd \sigma_0}=\left(1+\frac{\alpha}{2\pi}L{\cal{P}}^{(1)}_{\Delta}
+\frac{1}{2!}\left(\frac{\alpha}{2\pi}L\right)^2{\cal{P}}^{(2)}_{\Delta}
\right)^4,\nonumber \\
&&{\cal{P}}^{(1)}_{\Delta}=2\ln\Delta+\frac{3}{2}, \qquad
{\cal{P}}^{(2)}_{\Delta}=\left(2\ln\Delta+\frac{3}{2}\right)^2-4\frac{\pi^2}{6},
\qquad \Delta=\frac{\Delta\varepsilon}{\varepsilon}\, ,
\end{eqnarray}
and somewhat rewrite the main result of this paper:
\begin{eqnarray}\label{eq:21}
\Delta_{SS}+\Delta_{SV}+\sum\limits_{i=1}^3\Delta_i&=&
F^2L^2\left[
\frac{1}{2}{\cal{P}}^{(2)}_{\Delta}+\frac{3}{2}{\cal{P}}^{(1)}_{\Delta}\right]
+F^2\cdot\frac{4}{3}\biggl[L_{\lambda}L^2-L^3\biggr]
+L_{\lambda}L\cdot[F^2a_1+c_1] \nonumber \\
&+&L^2\biggl[F^2\left(a_2+2\frac{\pi^2}{6}-\frac{9}{2}\right)
+c_2\biggr]+L\cdot\biggl[F^2(b+z_1+z_2)+d\biggr],
\end{eqnarray}
Then, one can immediately be convinced that indeed an agreement takes place.
Besides, the non--leading terms of the types
$L\ln^2(\Delta\varepsilon/\varepsilon), L\ln(\Delta\varepsilon/\varepsilon)$
are one of the new incomings obtained in this paper.
We expect the 2--boxes contribution to compensate second, third and fourth
terms on rhs of Eq.~(\ref{eq:21}) and does modify the fifth one.

\subsection*{Acknowledgments}

We are grateful to N.~Merenkov, L.~Trentadue for discussions.
The work was supported in part by the INTAS foundation, grant 93--1867 ext.

\section*{Appendix A. Set of integrals corresponding to the
box--type Feynman diagrams}
\setcounter{equation}{0}
\renewcommand{\theequation}{A.\arabic{equation}}

We consider the large--angle high--energy Bhabha elastic scattering process.
An accuracy of formul\ae given below is predetermined by the omitted terms of
order of $m^2/s$ as compared to those of order unity.

The integrals for the scattering--type box diagram Fig.~2(m)
with uncrossed photon lines are (see also~\cite{prep})
\begin{eqnarray}
&&\int\frac{\dd^4k}{\mathrm{i}\pi^2}\frac{\{1,\ k^{\mu},\ k^{\mu}k^{\nu}\}}
{A(p_1+k)A(k)A(k-q)A(p_2-k)} = \biggl\{ G,
\quad G_1p_-^{\mu}+G_2q^{\mu}\,, \nonumber \\
&&\quad G_{00}g^{\mu\nu}+G_{11}p_-^{\mu}p_-^{\nu}+G_{22}q^{\mu}q^{\nu}
+G_{33}p_+^{\mu}p_+^{\nu}+G_{12}
(p_-^{\mu}q^{\nu}+p_-^{\nu}q^{\mu})\biggr\}, \\
&&q=p_1'-p_1=p_2-p_2'\,, \qquad p_{\pm}=\frac{1}{2}(p_2\pm p_1). \nonumber
\end{eqnarray}
Coefficients in the above expression look as follows:
\begin{eqnarray}
&&G=\frac{2}{st}L_s(L_t-L_{\lambda})\,,\quad G_1=-\frac{2}{us}L_s(L_t-L_s)
-\frac{2}{u}(\psi_s+\psi_t)\,,\quad G_2=\frac{1}{2}(G-G_1), \nonumber \\
&&G_{00}=-\frac{s}{4}G_1+\frac{1}{2}\psi_t\,,\quad G_{11}=\frac{s-t}{u}G_1
-\frac{4}{u}\psi_t+\frac{4}{tu}L_t+\frac{4}{us}L_s\,,\nonumber \\
&&G_{22}=\frac{s-u}{2u}G_1+\frac{1}{2}G+\frac{t-u}{stu}L_s-\frac{1}{u}\psi_t
+\frac{1}{tu}L_t\,,\quad G_{33}=G_1-\frac{4}{st}L_t\,, \nonumber \\
&&\quad G_{12}=-\frac{s}{u}G_1+\frac{2}{u}\psi_t-\frac{2}{tu}L_t
-\frac{2}{us}L_s\,,\quad \psi_t=\frac{1}{t}\left(\frac{1}{2}L_t^2
+\frac{2}{3}\pi^2\right)\,,\nonumber \\
&& \psi_s=\frac{1}{s}\left(\frac{1}{2}L^2+\frac{2}{3}\pi^2\right)\,,\quad
L_s=L-\mathrm{i}\pi.
\end{eqnarray}
The integrals for the scattering--type box diagram Fig.~2(n)
with crossed photon lines are
\begin{eqnarray}
&&\int\frac{\dd^4k}{\mathrm{i}\pi^2}\frac{\{1,\ k^{\mu},\ k^{\mu}k^{\nu}\}}
{A(p_1+k)A(k)A(k-q)A(p_2'+k)}=\biggl\{\widetilde G,
\quad \widetilde G_1\widetilde p_-^{\mu}+\widetilde G_2q^{\mu}\,, \nonumber \\
&&\widetilde G_{00}g^{\mu\nu}+\widetilde G_{11}p_-^{\mu}
p_-^{\nu}+\widetilde G_{22}q^{\mu}q^{\nu}
+\widetilde G_{33}\widetilde p_+^{\mu}\widetilde p_+^{\nu}+\widetilde G_{12}
(\widetilde p_-^{\mu}q^{\nu}+\widetilde p_-^{\nu}q^{\mu})\biggr\}, \\
&&q=p_1'-p_1=p_2-p_2'\,, \qquad \widetilde p_{\pm}=\frac{1}{2}(-p_2'\pm p_1).
\nonumber
\end{eqnarray}
where
\begin{eqnarray}
&&\widetilde G=\frac{2}{tu}L_u(L_t-L_{\lambda})\,,\quad \widetilde G_1
=-\frac{2}{us}L_u(L_t-L_u)
-\frac{2}{s}(\psi_u+\psi_t)\,,\quad \widetilde G_2=\frac{1}{2}(\widetilde G
-\widetilde G_1), \nonumber \\
&&\widetilde G_{00}=-\frac{u}{4}\widetilde G_1+\frac{1}{2}\psi_t\,,\quad
\widetilde G_{11}=\frac{u-t}{s}\widetilde G_1
-\frac{4}{s}\psi_t+\frac{4}{st}L_t+\frac{4}{us}L_u\,,\nonumber \\
&&\widetilde G_{22}=\frac{u-s}{2s}\widetilde G_1+\frac{1}{2}\widetilde G
+\frac{t-s}{stu}L_u-\frac{1}{s}\psi_t
+\frac{1}{st}L_t\,,\quad \widetilde G_{33}=\widetilde G_1-\frac{4}{tu}L_t\,, \nonumber \\
&&\quad \widetilde G_{12}=-\frac{u}{s}\widetilde G_1+\frac{2}{s}\psi_t-\frac{2}{st}
L_t-\frac{2}{us}L_u\,,\quad \psi_u=\frac{1}{u}\left(\frac{1}{2}L_u^2
+\frac{\pi^2}{6}\right)\,,\quad L_u=\ln\frac{-u}{m^2}.
\end{eqnarray}
The integrals for the annihilation--type box diagram Fig.~2(p)
with uncrossed photon lines are
\begin{eqnarray}
&&\int\frac{\dd^4k}{\mathrm{i}\pi^2}\frac{\{1,\ k^{\mu},\ k^{\mu}k^{\nu}\}}
{A(p_1+k)A(k)A(k-q_1)A(p_1'+k)}=\biggl\{ H,
\quad H_1p_-'^{\mu}+H_2q_1^{\mu}\,, \nonumber \\
&&H_{00}g^{\mu\nu}+H_{11}p_-'^{\mu}p_-'^{\nu}+H_{22}q_1^{\mu}q_1^{\nu}
+H_{33}p_+'^{\mu}p_+'^{\nu}+H_{12}
(p_-'^{\mu}q_1^{\nu}+p_-'^{\nu}q_1^{\mu})\biggr\}, \\
&&q_1=-p_1-p_2\,, \qquad p_{\pm}'=\frac{1}{2}(-p_1'\pm p_1). \nonumber
\end{eqnarray}
Coefficients are:
\begin{eqnarray}
&&H=\frac{2}{st}L_t(L_s-L_{\lambda})\,,\quad H_1=-\frac{2}{ut}L_t(L_s-L_t)
-\frac{2}{u}(\psi_{1s}+\psi_t)\,,\quad H_2=\frac{1}{2}(H-H_1), \nonumber \\
&&H_{00}=-\frac{t}{4}H_1+\frac{1}{2}\psi_{1s}\,,\quad H_{11}=\frac{t-s}{u}H_1
-\frac{4}{u}\psi_{1s}+\frac{4}{us}L_s+\frac{4}{ut}L_t\,,\nonumber \\
&&H_{22}=\frac{t-u}{2u}H_1+\frac{1}{2}H+\frac{s-u}{stu}L_t-\frac{1}{u}
\psi_{1s}+\frac{1}{su}L_s\,,\quad H_{33}=H_1-\frac{4}{st}L_s\,, \nonumber \\
&&\quad H_{12}=-\frac{t}{u}H_1+\frac{2}{u}\psi_{1s}-\frac{2}{su}L_s
-\frac{2}{tu}L_t\,,\quad \psi_{1s}=\frac{1}{s}\left(\frac{1}{2}L^2
+\frac{\pi^2}{6}-\mathrm{i}\pi L\right)\,,\\
&&L_s=L-\mathrm{i}\pi.\nonumber
\end{eqnarray}
The integrals for the annihilation--type box diagram Fig.~2(q)
with crossed photon lines are
\begin{eqnarray}
&&\int\frac{\dd^4k}{\mathrm{i}\pi^2}\frac{\{1,\ k^{\mu},
\ k^{\mu}k^{\nu}\}}{A(p_1+k)A(k)A(k-q_1)A(p_2'+k)}=\biggl\{\widetilde H,
\quad \widetilde H_1\widetilde p_-^{\mu}+\widetilde H_2q_1^{\mu}\,, \nonumber \\
&&\widetilde H_{00}g^{\mu\nu}+\widetilde H_{11}\widetilde p_-^{\mu}\widetilde p_-^{\nu}
+\widetilde H_{22}q_1^{\mu}q_1^{\nu}
+\widetilde H_{33}\widetilde p_+^{\mu}\widetilde p_+^{\nu}+\widetilde H_{12}
(\widetilde p_-^{\mu}q_1^{\nu}+\widetilde p_-^{\nu}q_1^{\mu})\biggr\}, \\
&&q_1=-p_1-p_2\,, \qquad \widetilde p_{\pm}=\frac{1}{2}(-p_1'\pm p_1). \nonumber
\end{eqnarray}
And the corresponding coefficients have the form
\begin{eqnarray}
&&\widetilde H=\frac{2}{su}L_u(L_s-L_{\lambda})\,,\quad \widetilde H_1=
-\frac{2}{ut}L_u(L_s-L_u)
-\frac{2}{t}(\psi_{1s}+\psi_u)\,,\quad \widetilde H_2=\frac{1}{2}
(\widetilde H-\widetilde H_1), \nonumber \\
&&\widetilde H_{00}=-\frac{u}{4}\widetilde H_1+\frac{1}{2}\psi_{1s}\,,\quad
\widetilde H_{11}=\frac{u-s}{t}\widetilde H_1
-\frac{4}{t}\psi_{1s}+\frac{4}{ts}L_s+\frac{4}{ut}L_u\,,\nonumber \\
&&\widetilde H_{22}=\frac{u-t}{2t}\widetilde H_1+\frac{1}{2}\widetilde H+\frac{s-t}
{stu}L_u-\frac{1}{t}
\psi_{1s}+\frac{1}{st}L_s\,,\quad \widetilde H_{33}=\widetilde H_1-\frac{4}{su}
L_s\,, \nonumber \\
&&\quad \widetilde H_{12}=-\frac{u}{t}\widetilde H_1+\frac{2}{t}\psi_{1s}
-\frac{2}{st}L_s-\frac{2}{tu}L_u.
\end{eqnarray}

\section*{Appendix B. Two soft photons emission}
\setcounter{equation}{0}
\renewcommand{\theequation}{B.\arabic{equation}}

In this appendix we give the explicit expressions of integrals
which can be encounted in considering a contribution coming
from two soft photons emission.
We have to calculate the following expression:
\begin{eqnarray}
\frac{\dd \sigma^{SS}}{\dd \sigma_0}&=&\frac{1}{2!}\left(\frac{-4\pi\alpha}{16\pi^3}
\right)^2
\int\frac{\dd^3\vecc{k}_1}{\omega_1}\int\frac{\dd^3\vecc{k}_2}{\omega_2}
\left(\frac{p_1'}{p_1'k_1}-\frac{p_1}{p_1k_1}+\frac{p_2}{p_2k_1}-\frac{p_2'}
{p_2'k_1}\right)^2 \nonumber\\
&\times&\left(\frac{p_1'}{p_1'k_2}-\frac{p_1}{p_1k_2}+\frac{p_2}{p_2k_2}
-\frac{p_2'}{p_2'k_2}\right)^2.
\end{eqnarray}
The region of integration over the energies of photons
obeys the strict inequality ${\cal D}:\ \omega_1+\omega_2 < \Delta E$.
The above formula when elaborated has three different structures.
The first one looks as follows:
\begin{eqnarray}
&&m^2\int\frac{\dd^3 \vecc{k}_1}{\omega_1}\left(\frac{1}{(p_1'k_1)^2}
+\frac{1}{(p_1k_1)^2}+\frac{1}{(p_2k_1)^2}+\frac{1}{(p_2'k_1)^2}\right)
\nonumber\\
&&\qquad = 16\pi\int\limits_{{\cal{D}}'} \sigma^2
\frac{x_1^2\dd x_1}{\sqrt{1+x_1^2}}
\left(1+\sigma^2 x_1^2\right)^{-1}
\equiv 16\pi\int\limits_{{\cal{D}}'}\dd x_1 f_1(x_1).
\end{eqnarray}
Here $x_1=k_1/\lambda$, $\sigma^2=m^2/\varepsilon^2$
and the region of integration transforms into
$${\cal D}':\quad 0 < x_1+x_2 < N=\Delta E/\lambda.$$
The second reads:
\begin{eqnarray}
&&\int\frac{\dd^3 \vecc{k}_1}{\omega_1}\left(\frac{2p_1p_2}
{p_1k_1\cdot p_2k_1}+\frac{2p_1'p_2'}{p_1'k_1\cdot p_2'k_1}\right)\\ \nonumber
&&\qquad = 16\pi\int\limits_{{\cal{D}}'}\frac{x_1\dd x_1}{1+x_1^2}\ln\frac{(
\sqrt{x_1^2+1}+x_1)^2}{\sigma^2 x_1^2+1}
\equiv 16\pi\int\limits_{{\cal{D}}'}\dd x_1 f_2(x_1),
\end{eqnarray}
and the third one is
\begin{eqnarray}
&&\!\!\!\!\int\frac{\dd^3 \vecc{k}_1}{\omega_1}\left(\frac{2p_1'p_2}
{p_1'k_1\cdot p_2k_1}+\frac{2p_1p_2'}{p_1k_1\cdot p_2'k_1}
-\frac{2p_1p_1'}{p_1k_1\cdot p_1'k_1}-\frac{2p_2p_2'}
{p_2k_1\cdot p_2'k_1}\right) \\
&&\hspace{.5cm}=16\pi\int\limits_{{\cal{D}}'}\frac{x_1\dd x_1}
{\sqrt{1+x_1^2}}\Biggl\{
\frac{\sqrt{\frac{1+c}{2}}}{\sqrt{1+x_1^2\frac{1+c}{2}}}
\ln\frac{\left(\sqrt{x_1^2\frac{1+c}{2}+1}+x_1\sqrt{\frac{1+c}{2}}\right)^2}
{\sigma^2 x_1^2+1} \nonumber \\ \nonumber
&&\hspace{.5cm}-\frac{\sqrt{\frac{1-c}{2}}}{\sqrt{1+x_1^2\frac{1-c}{2}}}
\ln\frac{\left(\sqrt{x_1^2\frac{1-c}{2}+1}+x_1\sqrt{\frac{1-c}{2}}\right)^2}
{\sigma^2 x_1^2+1} \Biggr\}
\equiv 16\pi\int\limits_{{\cal{D}}'}\dd x_1 [f_3^u(x_1)-f_3^t(x_1)].
\end{eqnarray}
Further evaluation gives for the terms of $(1)\ast (1)$ type:
\begin{eqnarray}
\int\!\!\int\limits_{\!\!\!\!\!{\cal D}'}\dd^2 xf_1(x_1)f_1(x_2)=
\left(\ln(2N)-\frac{1}{2}L\right)^2-\frac{\pi^2}{6}.
\end{eqnarray}
For those of $(1)\ast (2)$ type we have:
\begin{eqnarray}
\int\!\!\int\limits_{\!\!\!\!\!{\cal D}'}\dd^2 xf_1(x_1)f_2(x_2)
=-L\frac{\pi^2}{6}+\left(\ln(2N)-\frac{1}{2}L\right)
\left[L\left(\ln(2N)-\frac{1}{2}L\right)+\frac{1}{4}L^2-\frac{\pi^2}{6}\right].
\end{eqnarray}
And the remaining looks as follows:
\begin{eqnarray}
(1)\ast (3):
\int\!\!\int\limits_{\!\!\!\!\!{\cal D}'}\dd^2 xf_1(x_1)f_3^u(x_2)
 &=&\left(\ln(2N)-\frac{1}{2}L\right)
\biggl[L_u\left(\ln(2N)-\frac{1}{2}L\right) \nonumber\\
&+&\frac{1}{4}L_u^2-\frac{\pi^2}{6}+\Phi_u\biggr]-\frac{\pi^2}{6}L_u, \nonumber  \\
(2)\ast (2):
\int\!\!\int\limits_{\!\!\!\!\!{\cal D}'}\dd^2 xf_2(x_1)f_2(x_2)&=&
\left[L\left(\ln(2N)-\frac{1}{2}L\right)+\frac{1}{4}L^2
-\frac{\pi^2}{6}\right]^2-L^2\frac{\pi^2}{6}, \nonumber \\
(2)\ast (3):
\int\!\!\int\limits_{\!\!\!\!\!{\cal D}'}\dd^2 xf_2(x_1)f_3^u(x_2)&=&
\biggl[L\left(\ln(2N)-\frac{1}{2}L\right)+\frac{1}{4}L^2
-\frac{\pi^2}{6}\biggr] \\
&\times&\biggl[L_u\left(\ln(2N)-\frac{1}{2}L\right)+\frac{1}{4}L_u^2-\frac{\pi^2}{6}
+\Phi_u \biggr] - LL_u\frac{\pi^2}{6}, \nonumber \\
(3)_i\ast (3)_j:
\int\!\!\int\limits_{\!\!\!\!\!{\cal D}'}\dd^2 xf_3^i(x_1)f_3^j(x_2)&=&
\biggl[L_i\left(\ln(2N)-\frac{1}{2}L\right)+\frac{1}{4}L_i^2-\frac{\pi^2}{6}
+\Phi_i \biggr] \nonumber \\
&\times&\biggl[L_j\left(\ln(2N)-\frac{1}{2}L\right)+\frac{1}{4}L_j^2-\frac{\pi^2}{6}
+\Phi_j \biggr] - L_iL_j\frac{\pi^2}{6}\, , \nonumber
\end{eqnarray}
\begin{eqnarray*}
i=u,t,\quad L_u=\ln\frac{-u}{m^2},\quad L_t=\ln\frac{-t}{m^2},\quad
\Phi_u=\Li\left(\frac{s+u}{s}\right),\quad
\Phi_t=\Li\left(\frac{s+t}{s}\right).
\end{eqnarray*}
Summing up all the expressions derived and making definite rearrangements
we arrive to the final result given above in~(\ref{SS}).

\end{document}